\documentclass[reprint,aps,prd,nobibnotes,nofootinbib,twocolumn]{revtex4-2} 

\usepackage{amsmath}
\usepackage{amssymb}
\usepackage{graphicx}
\usepackage{fullpage}
\usepackage{hyperref}

\usepackage{color}
\definecolor{darkblue}{rgb}{0,0,0.5}

\begin{document}

\title{Factorization for Collider Dataspace Correlators}%

\author{Andrew J.~Larkoski}%
\email{larkoski@aps.org}
\affiliation{American Physical Society, Hauppauge, New York 11788, USA}

%\author{et al}

\date{\today}%

\begin{abstract}
\noindent A metric on the space of collider physics data enables analysis of its geometrical properties, like dimensionality or curvature, as well as quantifying the density with which a finite, discrete ensemble of data samples the space.  We provide the first systematically-improvable precision calculations on this dataspace, presenting predictions resummed to next-to-leading logarithmic accuracy, using the Spectral Energy Mover's Distance (SEMD) as its metric.  This is accomplished by demonstration of factorization of soft and collinear contributions to the metric at leading power and renormalization group evolution of the single-scale functions that are present in the factorization theorem.  As applications of this general framework, we calculate the two-point correlator between pairs of jets on the dataspace, and the measure of the non-Gaussian fluctuations in a finite dataset.  For the non-Gaussianities, our calculations validate the existence of a universal structure that had been previously observed in simulated data.  As byproducts of this analysis, we also calculate the two-loop anomalous dimension of the SEMD metric and show that the original Energy Mover's Distance metric is identical to the SEMD through next-to-next-to-leading logarithmic accuracy.
\end{abstract}

\maketitle

\tableofcontents

\section{Introduction}

While detected and measured on the two-dimensional celestial sphere, collider physics event data lives in a much larger, extremely high dimensional space.  This abstract dataspace is closely related to the phase space of the momentum of the produced final state particles, but events that cannot even in principle be experimentally distinguished are identified.  As any physically-realizable detector apparatus has lower thresholds on both the detected energy of particles as well as their angular resolution, events are identified if their detectable energy flow is indistinguishable.  This eminently reasonable physical constraint implies that collider physics events that populate this abstract dataspace are in some sense representative members of conjugacy classes, and that the dataspace that they occupy is the union of arbitrary particle phase spaces with identical energy flow.

This approach to defining and subsequently desiring to understand this dataspace imposes mathematical constraints on it.  In particular, defining collider events by their energy flow is infrared and collinear (IRC) safe \cite{Sterman:1977wj,Ellis:1996mzs}, and subsequently enables analysis order-by-order in the coupling of the gauge theory of interest.  This motivated a construction of an IRC safe metric between collider events from which the geometry of the dataspace could be studied and quantified.  This initial metric, the Energy Mover's Distance \cite{Komiske:2019fks} then motivated a large number of other IRC safe metrics that emphasized different aspects of the dataspace manifold \cite{Mullin:2019mmh,CrispimRomao:2020ejk,Cai:2020vzx,Larkoski:2020thc,Tsan:2021brw,Cai:2021hnn,Kitouni:2022qyr,Alipour-Fard:2023prj,Larkoski:2023qnv,Davis:2023lxq,Ba:2023hix,Craig:2024rlv,Cai:2024xnt,Gambhir:2024ndc,Larkoski:2025clo,Cai:2025fyw}.

While significant effort has been devoted to classifying these metrics, or from them defining observables, all studies thus far have been limited to analysis on simulated data, with calculations performed at only the simplest, lowest perturbative order.  Without concrete and systematically-improvable predictions, conclusions drawn from simulation can fail to clearly disentangle honest physical effects on the dataspace from idiosyncrasies of the simulations.  Our goal in this paper will be to present such a systematically-improvable framework for calculations on the dataspace manifold, within the perturbation theory of quantum chromodynamics (QCD).  Specifically, we will focus our study on the dynamics of jets, high energy collimated streams of particles in QCD, and we will present several example calculations.  However, we believe that this broader, and in some ways meta, approach to analyzing collider events from the space on which they are drawn is vastly rich, and the results we present here are mere introductions to what can be learned from this perspective.

The central object in this program is the dataspace manifold ${\cal M}$ equipped with a metric, $d(\Pi,\Pi')$, which satisfies all the properties of a metric, between two points on particle phase space, $\Pi$, $\Pi'$.  With this, the local geometry about any point on the manifold can be calculated from the distribution of points a fixed distance $d$ away:
\begin{align}
&\int d\Pi'\, p(\Pi')\, \delta\left(d- d(\Pi,\Pi')\right)\\
&\hspace{1.5cm}\propto d^{\text{dim}(\Pi)-1}\left(1-\frac{{\cal S}(\Pi)}{6\cdot\text{dim}(\Pi)}\,d^2+\cdots\right)\,,
\nonumber
\end{align}
where $p(\Pi)$ is the probability distribution of events on phase space.  In the second line, we have expanded this distribution in the small distance limit, $d\to 0$.  The overall scaling is set by the dimension of the manifold about the point $\Pi$, $\text{dim}(\Pi)$, and the first corrections are proportional to the scalar curvature, ${\cal S}(\Pi)$.  The averaged dimensionality on this space has been studied in data \cite{Komiske:2019jim}, as defined through the correlation dimension \cite{Grassberger:1983zz}.  Information about the topology of the manifold can be defined through its persistent homology \cite{edelsbrunner2002topological,carlsson2009topology}, as represented as a simplicial complex.  Further, volumes of simplices formed from points on the manifold can be calculated with the Cayley-Menger determinant \cite{cayley1841theorem,menger1928untersuchungen}.

A central problem in collider physics, and jet physics in particular, is the classification or discrimination of events according to the features of their measured energy distribution.  This problem has been a fundamental driver of advances in jet substructure \cite{Seymour:1993mx,Butterworth:2008iy,Abdesselam:2010pt,Altheimer:2012mn,Altheimer:2013yza,Adams:2015hiv} and machine learning in collider physics, see, e.g., Refs.~\cite{Larkoski:2017jix,Kogler:2018hem,Guest:2018yhq,Albertsson:2018maf,Radovic:2018dip,Carleo:2019ptp,Bourilkov:2019yoi,Schwartz:2021ftp,Karagiorgi:2021ngt,Boehnlein:2021eym,Shanahan:2022ifi,Plehn:2022ftl,Nachman:2022emq,DeZoort:2023vrm,Zhou:2023pti,Belis:2023mqs,Mondal:2024nsa,Feickert:2021ajf,Larkoski:2024uoc,Halverson:2024hax} for reviews.  Within the problem of binary discrimination, for which signal and background classes are well-defined and whose probability distributions are known, the optimal discrimination observable is the likelihood ratio, by the Neyman-Pearson lemma \cite{Neyman:1933wgr}.  However, only given discrete and finite signal and background data ensembles, the likelihood ratio cannot be directly evaluated on the dataspace, but instead its functional form must be parametrized and subsequently optimized in some way.  With a metric, the discrete data can be smeared into a continuous distribution on the dataspace, and a corresponding smeared likelihood ratio defined \cite{Larkoski:2025clo}:
\begin{align}
{\cal L}(\Pi|d) = \frac{\int d\Pi'\, p_b(\Pi')\,\Theta\left(d-d(\Pi,\Pi')\right)}{\int d\Pi'\, p_s(\Pi')\,\Theta\left(d-d(\Pi,\Pi')\right)}\,,
\end{align}
where $d$ here is the smearing radius, and the subscripts $b$ and $s$ represent background and signal distributions on phase space, respectively.

Variation of the smearing radius $d$ modifies the sensitivity of the smeared likelihood to physics at different scales, and as $d\to 0$, the true likelihood is obtained.  On a finite dataset, however, there will be some minimal smearing radius $d_{\min}$ that ensures that there is always at least one event within a distance $d_{\min}$ of the phase space point of interest $\Pi$.  For a dataset of $N$ events, this minimal smearing radius is defined implicitly through
\begin{align}
\int d\Pi'\, p(\Pi')\, \Theta\left(d_{\min} - d(\Pi,\Pi')\right)= \frac{1}{N}\,.
\end{align}
The statistical properties of this minimal distance given a distribution on phase space $p(\Pi)$ and a data size $N$ can be analyzed through extreme value theory \cite{frechet1927loi,fisher1928limiting,von1936distribution,gnedenko1943distribution}, and smearing radius prescriptions can be made accordingly.

These examples demonstrate that a wide range of physics results are accessible with an event metric.  However, each is dependent on the choice of coordinates on phase space and subsequently dataspace, which greatly complicates possible calculations, as the relevant dimensionality of these spaces can be in the hundreds.  From a geometrical perspective, such a formulation would not be diffeomorphism invariant, and so concrete physical consequences can be challenging to interpret. Coordinate independence and diffeomorphism invariance can be easily solved by considering moments, or $n$-point event correlation functions, of events on the dataspace.  The general $n$-point correlator between all ${n\choose 2}$ pairs would be defined as
\begin{widetext}
\begin{align}
p_n\left(
\{d_{ij}\}_{i<j}^n
\right)\equiv \int \left[
\prod_{a=1}^n d\Pi_a\,p(\Pi_a)\, 
\right]\prod_{i<j}^n \delta\left(
d_{ij} - d(\Pi_i,\Pi_j)
\right)\,.
\end{align}
\end{widetext}
In this form, these $n$-point event correlators are similar in spirit to multi-point energy correlators, e.g., Refs.~\cite{Basham:1978bw,Hofman:2008ar,Dixon:2018qgp,Chen:2020vvp}.  However, energy correlators are measured between pairs of points on the celestial sphere in a single event, weighted by the flow of energy at those points, and averaged over the event ensemble.

It turns out that the two-point event correlator as a function of distance $d$ is sufficient to completely determine the dataspace manifold \cite{boutin2004reconstructing}, with probability 1.  However, successfully inverting the information of the two-point correlator is challenging, as it is related to inverting the distance matrix between all pairs of $N$ events sampled on the manifold.  Higher-point correlators provide significantly more information without the necessity of (extremely high dimensional) matrix inversion.  Restricting to information in the two- and three-point event correlators, symmetries of the dataspace manifold under translation can be established \cite{Larkoski:2025idq}.

In this paper, we present and construct a systematically-improvable framework for calculation of the $n$-point event correlators.  Explicit calculation requires a choice of metric, and we use the Spectral Energy Mover's Distance (SEMD) \cite{Larkoski:2023qnv,Gambhir:2024ndc}, as it is currently the only IRC safe event metric that can be expressed exactly in closed-form, is invariant to event isometries, and can be evaluated on (simulated) data hundreds or thousands of times faster than other metrics.  Our calculations will focus on the small distance limit, $d\to 0$, in which the emissions that dominate the metric are either soft or collinear in origin.  To leading power in the SEMD metric distance $d$, we will prove that soft and collinear contributions factorize from one another to all orders in perturbation theory.  This enables a formulation of a factorization theorem in which logarithms of distance $d$ can be resummed through renormalization group evolution of the functions in the factorization theorem.

This paper is organized as follows.  In Sec.~\ref{sec:semd}, we define the SEMD and discuss its properties as a metric on collider event dataspace.  In Sec.~\ref{sec:powercount}, we perform the power counting analysis of the SEMD to explicitly demonstrate that soft and collinear contributions factorize at leading power.  This enables us in Sec.~\ref{sec:factth} to construct and present the factorization theorem for the SEMD event correlators.  Here, we will focus on the two-point correlator, but also show how the expressions generalize to higher-point correlators.  In Sec.~\ref{sec:nongauss}, we use the factorization theorem to calculate the two- and three-point correlators to next-to-leading logarithmic (NLL) accuracy.  For the two-point correlator, the extra Sudakov suppression from sampling two events on the dataspace actually renders the distribution perturbatively calculable over almost its entire domain of support.  With the three-point correlator, we calculate a measure of non-Gaussianity and provide validating evidence for the universality of structure that was first observed in simulated events in Ref.~\cite{Larkoski:2025idq}.  We conclude in Sec.~\ref{sec:concs}, and look forward to further precision calculations and ultimately comparison with experimental data.

Calculational details are presented in the appendices.  App.~\ref{app:nllresum} presents all necessary details for accomplishing resummation through NLL accuracy.  Significant discussion is devoted to performing the inverse Laplace transformation of the three-point event correlator, as no exact expression is known, so we must resort to approximations.  In App.~\ref{app:twoloop}, we present the explicit calculation of the two-loop anomalous dimension of the two-point correlator's jet function.  This is actually very closely related to the two-loop anomalous dimension for the (regular) jet mass, modified by a contribution that correlates two events, each at one-loop accuracy.  While we do not use this result in this paper, all pieces are known for resummation of the two-point event correlator through next-to-next-to-leading logarithmic accuracy (NNLL).  Finally, in App.~\ref{app:temdrg}, we compare the SEMD to the expression for the (original) EMD evaluated between two jets, each with two particles in them, first derived in Ref.~\cite{Larkoski:2023qnv}. We demonstrate that these expressions are identical in the appropriate limits (up to an overall factor), proving that the SEMD and the EMD are identical through NNLL accuracy.

%\begin{align}
%&p\left({\cal P}(d)\right)=\\
%&\int d\Pi\, p(\Pi)\, \delta\left(
%{\cal P}(d) - \int d\Pi'\, p(\Pi')\, \delta\left(d - d(\Pi,\Pi')\right)
%\right)\nonumber
%\end{align}
%
%\begin{align}
%&\langle {\cal P}^n(d)\rangle \\
%&\hspace{0.5cm}= \int d\Pi\, p(\Pi)\, \left(
%\int d\Pi'\, p(\Pi')\, \delta\left(d - d(\Pi,\Pi')\right)
%\right)^n\nonumber
%\end{align}

\section{Spectral Energy Mover's Distance Metric}\label{sec:semd}

In this section, we review the Spectral Energy Mover's Distance (SEMD) metric \cite{Larkoski:2023qnv,Gambhir:2024ndc} that we will use throughout the rest of this paper.  We will specifically focus on the $p=2$ form of the metric, as it is the currently only known particle physics collider event data metric that can be expressed exactly in closed form (with no residual implicit minimizations), is invariant to obvious event isometries, is infrared and collinear safe, and evaluates extremely fast, and so can be evaluated on large (simulated) datasets.

The general expression for the $p=2$ SEMD metric between the energy distributions of events $A$ and $B$ is
\begin{widetext}
\begin{align}\label{eq:semdtot}
\text{SEMD}_{p=2}({\cal E}_A,{\cal E}_B) = \sum_{i<j\in{\cal E}_A}2E_jE_i\omega_{ij}^2+\sum_{i<j\in{\cal E}_B}2E_iE_j\omega_{ij}^2 - 2\sum_{\substack{n\in{\cal E}_A^2\,, \ell\in{\cal E}^2_B\\\omega_n<\omega_{n+1}\\\omega_\ell<\omega_{\ell+1}}}\omega_n\omega_\ell \,\text{ReLU}(S_{n\ell})\,.
\end{align}
\end{widetext}
In this expression, $i$ and $j$ label individual particles in the events, and $\omega_{ij}$ is the appropriately-defined angle between the momenta of those particles.  In the term at furthest right, the subscripts $n,\ell$ labels pairs of particles from event $A$ and $B$, respectively, and the angles are ordered; $\omega_n < \omega_{n+1}$, for example.  The function $\text{ReLU}(x) = x\,\Theta(x)$, where $\Theta(x)$ is the Heaviside step function, is the rectified linear unit function \cite{4082265}.  Its argument in the SEMD is
\begin{align}
&S_{n\ell}\\
 &= \min\left[
S_A^+(\omega_n),S_B^+(\omega_\ell)
\right] - \max\left[
S_A^-(\omega_n),S_B^-(\omega_\ell)
\right]\,.\nonumber
\end{align}
The inclusive and exclusive cumulative spectral functions are
\begin{align}
S^+(\omega_n) &=\sum_{i\in{\cal E}}E_i^2+\sum_{\substack{n\geq m\in{\cal E}^2\\\omega_m<\omega_{m+1}}}(2EE)_m\,,\\
S^-(\omega_n) &=\sum_{i\in{\cal E}}E_i^2+\sum_{\substack{n> m\in{\cal E}^2\\\omega_m<\omega_{m+1}}}(2EE)_m\,,
\end{align}
where we have used the shorthand for the pair of particles in an event, $m=(i,j)$.

The SEMD is a metric, and as such, satisfies the properties of a metric.  Two are obvious from its definition above:
\begin{itemize}
\item Non-Negativity: $\text{SEMD}_{p=2}({\cal E}_A,{\cal E}_B)\geq 0$
\item Symmetry:\\$\text{SEMD}_{p=2}({\cal E}_A,{\cal E}_B)=\text{SEMD}_{p=2}({\cal E}_B,{\cal E}_A)$
\end{itemize}
The isometry invariance of the SEMD is accomplished through exclusive sensitivity to pairwise particle angles, with no reference to absolute collider detector coordinates.  As such, the identity of indiscernibles holds for almost all pairs of events; that is, it holds with probability 1:
\begin{itemize}
\item Identity of Indiscernibles:\\
$\text{SEMD}_{p=2}({\cal E}_A,{\cal E}_B)=0\leftrightarrow {\cal E}_A={\cal E}_B$
\end{itemize}
Finally, the SEMD is actually a measure of the squared metric distance between events, and so the triangle inequality holds for the square-roots of the SEMD:
\begin{itemize}
\item Triangle Inequality:\\
\begin{align*}
&\sqrt{\text{SEMD}_{p=2}({\cal E}_A,{\cal E}_B)}+\sqrt{\text{SEMD}_{p=2}({\cal E}_B,{\cal E}_C)}\\
&\hspace{4cm}\geq \sqrt{\text{SEMD}_{p=2}({\cal E}_A,{\cal E}_C)}
\end{align*}
\end{itemize}

For compactness in equations that follow, we will typically drop the $p=2$ subscript, as this will be the only metric we consider in the body of this paper.  However, in App.~\ref{app:temdrg}, we will present a simple comparison with the (tangent) Energy Mover's Distance metric \cite{Komiske:2019fks,Komiske:2020qhg}.  In particular, we will show that all quantities evaluated on the SEMD and EMD are identical through NNLL accuracy.  Therefore, while we will exclusively present results and arguments in the paper for the SEMD, they will hold for the original EMD, with necessary minor rescaling.

For all calculations that follow in this paper, we will choose the angular distance $\omega_{ij}$ to be
\begin{align}
\omega_{ij}^2 = 1-\cos\theta_{ij}\,.
\end{align}
In general, all that is required of $\omega_{ij}$ is that it is monotonically-increasing with true angle $\theta_{ij}$, but this particular choice makes the first two terms of Eq.~\ref{eq:semdtot} the total invariant masses of events $A$ and $B$.  With this particular choice, we can write down the expression for the SEMD as evaluated between two jets $A$ and $B$, where each jet consists of two particles.  The phase space of a jet with two particles can be expressed by its invariant mass $s$ and one particle's energy fraction $z$, and we assume rotational symmetry about the axis of the jet.  The SEMD between these two jets $A$ and $B$ is
\begin{align}\label{eq:semdtwopart}
&\text{SEMD}(\{s_A,z_A\},\{s_B,z_B\}) \\
&= s_A+s_B-2\sqrt{s_As_B}\sqrt{\frac{\min[z_A(1-z_A),z_B(1-z_B)]}{\max[z_A(1-z_A),z_B(1-z_B)]}}\,.\nonumber
\end{align}
If the emission in jet $B$ for example, becomes unresolved, either $z_B\to 0,1$ or $s_B\to 0$, the SEMD reduces to the invariant mass of jet $A$.  This expression and its simple limit will be exploited throughout the expressions in this paper.  Importantly, the close relationship of the SEMD to the mass of a jet means that almost no new calculations need to be performed, once we establish the contributions to the SEMD in appropriate limits.

\section{Power Counting the Two-Point Correlation}\label{sec:powercount}

Given the form of the SEMD metric between two arbitrary events or jets, our goal will be to calculate the distribution of pairwise distances.  As discussed in the Introduction, our central quantity of interest is the pairwise distance distribution or two-point event correlation,
\begin{align}
&p(d^2)\\
&\!=\int d\Pi_A\, d\Pi_B\, p(\Pi_A)\,p(\Pi_B)\,\delta\left(d^2-\text{SEMD}(\Pi_A,\Pi_B)\right)\,,\nonumber
\end{align}
where $p(\Pi)$ is the probability distribution of events on particle phase space $\Pi$.  In this section, we will establish the scaling behavior of the SEMD, and correspondingly this distribution, to all orders in the coupling.  This analysis will then enable us to write down an all-orders factorization theorem in the following section and from it, resum the large logarithms of the distance $d$ that arise to whatever accuracy we are strong enough to calculate.

We will restrict our analysis to the measurement and calculation of the SEMD distance between pairs of jets with a common total energy $E$.   Further, we make the assumption that distances $ d $ are parametrically small compared to energy $E$, $ d \ll E$.  With these restrictions, we then focus on identification of the dominant or leading-power contributions in this limit, to arbitrary orders in the coupling.  Because of the IRC safety of the SEMD and the soft and collinear divergences in fixed-order matrix elements, the dominant contributions to the two-point correlator in this limit come from emissions that are hard and collinear to the jet axis or low energy, but emitted at relatively wide angles.  Further, we know that squared matrix elements factorize in these soft or collinear limits \cite{Low:1958sn,Weinberg:1965nx,Burnett:1967km,Gribov:1972ri,Gribov:1972rt,Lipatov:1974qm,Dokshitzer:1977sg,Altarelli:1977zs}, and so to demonstrate factorization of the two-point correlator, we need to demonstrate that the soft and collinear contributions to the SEMD also factorize from one another, at least to leading power in the $ d \ll E$ limit.

We will accomplish this through power counting the possible contributions from the individual jets to the SEMD.  This power counting in the soft and collinear limits establishes the dominant modes in soft-collinear effective theory (SCET) \cite{Bauer:2000ew,Bauer:2000yr,Bauer:2001ct,Bauer:2001yt}, and in which matrix elements of operators can be renormalized to resum large logarithms.  Our analysis will not need the details and machinery of SCET specifically, because we will be able to recycle many results from the literature, e.g., Refs.~\cite{Schwartz:2007ib,Stewart:2010tn,Ellis:2010rwa,Larkoski:2014gra} (see also Ref.~\cite{Banfi:2004yd}), due to the close relationship between the SEMD and the jet mass.  In particular, our measurement of the distance $ d \ll E$ sets the relevant scale of the contributions.  To track this scaling, we will introduce the parameter $\lambda\ll 1$, where $ d  = \lambda E$, and expand the soft or collinear contributions to the SEMD to leading non-trivial order in $\lambda$.  We will consider all possible pairwise contributions: collinear-with-collinear, soft-with-soft, and collinear-with-soft, and this will inform the structure of the factorization theorem.

\subsection{Jet-Jet Correlations}

We first consider contributions to the SEMD from collinear emissions from both jets $A$ and $B$.  We define collinear emissions as those for which their pairwise opening angle is small, $\theta_{cc} \ll 1$, but their energy fraction is large, $E_c\sim E$.  In this limit, we can expand each term in the expression for the SEMD to leading order in the collinear limit for both jets.  We have
\begin{align}
\text{SEMD}&\xrightarrow{\text{coll-coll}} \sum_{A}(EE)_{c,A}\theta_{cc,A}^2+\sum_{B}(EE)_{c,B}\theta_{cc,B}^2\nonumber\\
&\hspace{1cm}-\sum_{A,B}\theta_{cc,A} \theta_{cc,B}\,\text{ReLU}(S_{c,AB})\,.
\end{align}
For compactness, we have suppressed the ordering of angles in the subtraction term.  In this term, the ReLU function is exclusively dependent on collinear particle energy.  By assumption, collinear particles have order-1 energy, and so the ReLU function too will have order-1 scaling in the collinear limit.

To compare terms amongst each other, we now rescale all collinear angles by $\lambda$, $\theta_{cc}\to \lambda\theta_{cc}$, and take only the leading power terms.  In this case, it is obvious that the SEMD is homogeneous in the collinear limit,
\begin{align}
\text{SEMD}&\xrightarrow{\theta_{cc}\to \lambda\theta_{cc}} \lambda^2\, \text{SEMD}\,,
\end{align}
and thus all terms contribute at leading power in this limit.

\subsection{Soft-Soft Correlations}

Now, let's move on to correlations between soft emissions in both jets $A$ and $B$.  A soft particle is defined as one for which its energy is parametrically small compared to the jet energy, $E_s\ll E$, but its angle from the jet axis, and from other soft particles, is order-1: $\theta_s\sim 1$.  In the evaluation of the contribution to the mass of a jet from soft particles, there are two possibilities.  Either a soft particle is correlated with hard, collinear particles, or it is correlated with other soft particles.  The contribution of the invariant mass of a jet from soft particles then takes the parametric form:
\begin{align}
m_s^2\sim \sum_{c,s} 2E_cE_s\omega_s^2 + \sum_{s,s'}2E_sE_{s'}\omega_{ss'}^2\,,
\end{align}
where $c$ denotes collinear particles and $s$ and $s'$ denote soft particles.  Because the collinear energy is order-1, $E_c\sim E$, and the soft angles are order-1, $\theta_{ss'}\sim 1$, the second term is parametrically smaller than the first, and so can be ignored to leading power.  That is, the dominant contribution to the squared jet mass from soft particles is
\begin{align}
m_s^2\sim \sum_{c,s} 2E_cE_s\omega_s^2\,.
\end{align}
Further at leading power, the sum over hard collinear particles can be done explicitly, because soft particles carry no energy and the angle between collinear particles cannot be resolved by the softs:
\begin{align}
m_s^2\sim 2E\sum_{s} E_s\omega_s^2\,.
\end{align}

Then, with this simplification, we expand the SEMD in the soft limit, where in each term, a soft particle is correlated with a hard particle.  We have
\begin{align}
\text{SEMD}&\xrightarrow{\text{soft-soft}} 2E\sum_{A}E_{s,A}\omega_{s,A}^2+2E\sum_{B}E_{s,B}\omega_{s,B}^2 \nonumber\\
&\hspace{1cm}- 2\sum_{A,B}\omega_{s,A}\omega_{s,B} \,\text{ReLU}(S_{s,AB})\,.
\end{align}
We still need to expand the ReLU term in the soft limit, but even without doing that, we can establish the necessary scaling behavior.  The ReLU term is quadratic in particle energies, so there are only 3 possibilities: it is independent of soft particle energies, it is linear in soft particle energies, or it is quadratic in soft particle energies.  If the ReLU term is quadratic in soft particle energies, then as argued earlier, this is subleading power, and can be ignored.  If it is linear in soft particle energy, then that is leading power, as the squared mass contributions are also linear in soft energies.

The more subtle case is the possibility that the ReLU term is independent of soft energies; or, that the ReLU term scales like order-1.  If this possibility were realized, then the soft contribution to the SEMD would be negative, as it would completely overwhelm the invariant mass contributions.  By the non-negativity of the SEMD as a metric, this is impossible.  Therefore, the ReLU term must scale linearly in the soft energy.

With the scaling we had already established from collinear correlations, the squared distance or invariant mass scales like $\lambda^2$.  This necessarily requires that the energy of soft emissions scales like $\lambda^2$, $E_s\sim \lambda^2$, for soft and collinear contributions to both be leading power.  That is, with the scaling $E_s\to \lambda^2 E_s$ for all soft particle energies, the SEMD in the soft limit is homogeneous:
\begin{align}
\text{SEMD}&\xrightarrow{E_{s}\to \lambda^2E_{s}} \lambda^2\, \text{SEMD}\,,
\end{align}
and thus all terms contribute at leading power in this limit.

\subsection{Jet-Soft Correlations}

Finally, we must consider the possibility that there are collinear emission contributions from one jet, and soft emission contributions from the other jet.  By the symmetry of the SEMD, we can without loss of generality, assume that collinear emissions come from jet $A$, and soft emissions from jet $B$.  Then, taking the leading-power soft and collinear limits appropriately, the SEMD takes the form
\begin{align}
\text{SEMD}&\xrightarrow{\text{soft-coll}} \sum_{A}(EE)_{c,A}\theta_{cc,A}^2+2E\sum_{B}E_{s,B}\omega_{s,B}^2 \nonumber\\
&\hspace{0cm}- \sqrt{2}\sum_{A,B}\theta_{cc,A}\omega_{s,B} \,\text{ReLU}(S_{cs,AB})\,.
\end{align}
As argued above, the collinear and soft invariant mass contributions are both at leading power, and scale like $\lambda^2$.

By contrast, the subtraction ReLU term is actually subleading power.  Note that there is an explicit collinear angle $\theta_{cc,A}$ in this term, and so it at least scales like $\lambda$.  As argued above, it is inconsistent for the energy dependence in the ReLU function to scale independently of $\lambda$, because that would render the SEMD negative, and violate general properties of the metric.  The only other possibility is that the ReLU function is linear in soft particle energies.  If that is true, then the whole subtraction term is at subleading power because the product of a collinear angle and soft energy scales like $\lambda^3$: $\theta_{cc}E_s\sim \lambda^3$.  Therefore, there is no soft-collinear mixing term at leading power.

This final property establishes the factorization of the SEMD into collinear and soft contributions at leading power in the $ d \ll E$ limit.  Further, these arguments used the scaling properties of the complete expression of the SEMD, with any number of particles in the jets, and therefore hold to all perturbative orders.  The expression for the SEMD to leading power in the soft and/or collinear limits is thus:
\begin{align}
\text{SEMD}&\xrightarrow{\text{soft,coll}} \sum_{A}(EE)_{c,A}\theta_{cc,A}^2+\sum_{B}(EE)_{c,B}\theta_{cc,B}^2\nonumber\\
&\hspace{1cm}+2E\sum_{A}E_{s,A}\omega_{s,A}^2+2E\sum_{B}E_{s,B}\omega_{s,B}^2 \nonumber\\\
&\hspace{1cm}-\sum_{A,B}\theta_{cc,A} \theta_{cc,B}\,\text{ReLU}(S_{c,AB})\nonumber\\
&\hspace{1cm}- 2\sum_{A,B}\omega_{s,A}\omega_{s,B} \,\text{ReLU}(S_{s,AB})\,.
\end{align}

\section{Factorization Theorem}\label{sec:factth}

With this power counting and factorization of the SEMD result, we are now able to state the factorization theorem for the two-point jet-jet correlator.  For concreteness, we will consider the SEMD measured between a jet in one collider event and another jet in another, independent, collider event.  The jets are assumed to be found with the same algorithm, and have the same energy $E$ (or relevant energy quantity, depending on the collider).  Between these two jets, we measure the SEMD, which returns some squared metric distance $ d ^2$.  We work to leading power in the limit $ d \ll E$, but to all orders in the QCD coupling $\alpha_s$.

With this set-up, we introduce the pairwise jet-jet and soft-soft correlations between two jets $A$ and $B$, as follows:
\begin{align}\label{eq:jetsoftdefs}
&J_{AB}( d ^2) \equiv\\
&\int d\Pi_A\, d\Pi_B\, J(\Pi_A)\, J(\Pi_B)\,\delta\left(
 d ^2 - \text{SEMD}(\Pi_A,\Pi_B)
\right)\,,\nonumber\\
&S_{AB}( d ^2) \equiv\\
&\int d\Pi_A\, d\Pi_B\, S(\Pi_A)\, S(\Pi_B)\,\delta\left(
 d ^2 - \text{SEMD}(\Pi_A,\Pi_B)
\right)\,.\nonumber
\end{align}
These quantities measure the distribution of the SEMD between collinear emissions in the jets, or soft emissions in the jets, as discussed in the previous section.  $\Pi$ is the appropriate phase space, $J(\Pi)$ is the jet function on that phase space that follows from collinear splitting functions, and $S(\Pi)$ is the squared soft current describing soft emissions from the dipoles throughout the event.

Then, the factorization theorem for the SEMD or squared distance distribution in the $ d ^2\ll E^2$ limit takes the form at leading power of
\begin{align}
&p( d ^2) \\
&=H_{AB} \int d (d _J^2)\, d (d _S^2)\, J_{AB}( d _J^2)\, S_{AB}( d _S^2)\,\delta( d ^2 -  d _J^2- d _S^2)\,,\nonumber
\end{align}
where $H_{AB}$ is the hard function that describes the hard collision event and subsequent jet production.  The hard function might still involve several disparate scales, especially at a hadron collider where parton distributions of the initial state need to be included, and further jet production may need to be suppressed.  Additionally, soft radiation may produce correlations between in- and out-of-jet scales resulting in non-global logarithms \cite{Dasgupta:2001sh}.  Thus, this expression may need to be further factorized to completely resum logarithms of all disparate scales.  We leave such an analysis to future work, and here focus on the novel properties introduced by the SEMD as a measure of correlations between events.

Specifically, unlike nearly all previous collider observables considered, this jet-jet or event-event distance is not measured exclusively on a single event, and then binned as a histogram or other distribution.  The di-event correlation of the SEMD, or other event metrics, means that the units or dimensionality of the corresponding distribution are rather unfamiliar.  A typical event shape distribution as measured in a collider experiment, would be expressed as a cross section, an area, possibly differential in the measured observable.  For the SEMD, even in this two-point correlator case, the distribution would be measured as a squared cross section, as there are two differential distributions on phase space, one for each event, that contribute.  In this way, this is in some sense similar to double parton scattering at a hadron collider, see, e.g. Ref.~\cite{Diehl:2017wew}.  However, double parton scattering is correlated, and possibly strongly so, as both sets of colliding partons are pulled from just two protons.  By contrast event-event correlations with a metric assumes independence between events from distinct collisions.

\subsection{Arbitrary-Point Correlations}

The soft and collinear factorization of the SEMD can be exploited to express the general $n$-point event distribution, in the limit that all pairwise distances are small compared to the jet's energy.  For example, the three-point event correlator distribution would be triply differential in each of the possible pairwise distances, where
\begin{align}
&p( d _{AB}^2, d _{AC}^2, d _{BC}^2)=H_{ABC}\\
&\times\int d (d _{AB,J}^2)\,d (d _{AC,J}^2)\,d (d _{BC,J}^2)\nonumber\\
&\times\int d (d _{AB,S}^2)\,d (d _{AC,S}^2)\,d (d _{BC,S}^2)\,\nonumber\\
&\hspace{2cm}\times J_{ABC}( d _{AB,J}^2, d _{AC,J}^2, d _{BC,J}^2)\nonumber\\
&\hspace{2cm}\times S_{ABC}( d _{AB,S}^2, d _{AC,S}^2, d _{BC,S}^2)\nonumber\\
&\hspace{2cm}\times \delta( d _{AB}^2- d _{AB,J}^2- d _{AB,S}^2)\nonumber\\
&\hspace{2cm}\times \delta( d _{AC}^2- d _{AC,J}^2- d _{AC,S}^2)\nonumber\\
&\hspace{2cm}\times \delta( d _{BC}^2- d _{BC,J}^2- d _{BC,S}^2)
\nonumber\,.
\end{align}
The hard function $H_{ABC}$ describes the three collision events and subsequent jet production, while the triply-differential jet function, for example, is defined as
\begin{align}
&J_{ABC}( d _{AB}^2, d _{AC}^2, d _{BC}^2) \\
&\hspace{1cm}\equiv\int d\Pi_A\, d\Pi_B\,d\Pi_C\, J(\Pi_A)\, J(\Pi_B)\,J(\Pi_C)\nonumber\\
&\hspace{2cm}\times\delta\left(
 d _{AB}^2 - \text{SEMD}(\Pi_A,\Pi_B)
\right)\nonumber\\
&\hspace{2cm}\times\delta\left(
 d _{AC}^2 - \text{SEMD}(\Pi_A,\Pi_C)
\right)\nonumber\\
&\hspace{2cm}\times\delta\left(
 d _{BC}^2 - \text{SEMD}(\Pi_B,\Pi_C)
\right)\,,\nonumber
\end{align}
and similarly for the soft function.

This factorization structure naturally generalizes to higher-point correlations.  For $n$-point correlations, there are of course ${n\choose 2}$ distinct pairwise distances that can be formed, and so the expression of the general factorization theorem grows extremely large.  As such, we will not write down the expression for the $n$-point event correlation factorization theorem in complete generality, but will instead focus on the two- and three-point correlators as exemplar.

\subsection{Structure of One-Loop Renormalization}\label{sec:oneloopresum}

The power of the factorization theorem is in the simplicity with which resummation of large logarithms is performed through renormalization group evolution.  To perform the factorization of the measurable distribution, we must introduce an unphysical scale, the renormalization scale $\mu$.  Each function in the factorization theorem is sensitive to a single physical, dimensionful scale, and so the argument of the functions must be the dimensionless ratio of the physical scale to the renormalization scale.  Demanding that the physical distribution is independent of the renormalization scale generates the renormalization group evolution and resums logarithms.

Focusing first on the two-point correlator, note that its factorized cross section is a convolution of functions:
\begin{align}
p( d ^2) &=H_{AB}(\mu,Q) \\
&\hspace{-0.8cm}\times \int d (d _J^2)\, d (d _S^2)\,J_{AB}(\mu, d _J^2)\, S_{AB}(\mu, d _S^2)\,\delta( d ^2 -  d _J^2- d _S^2)\,,\nonumber
\end{align}
because the form of the SEMD is as a sum of soft and collinear contributions, and we have included explicit dependence on the renormalization scale $\mu$.  Additionally, we have added the argument $Q$ to the hard function, which denotes whatever relevant hard scale $Q$ at which the jets are produced.  This will include the jet energy, $E$, but may include other scales, which is why we use the more general $Q$ here.  As a convolution for which the argument is non-negative, it is natural to Laplace transform in which the distribution takes the form of a simple product:
\begin{align}
\tilde p(\nu) &=H_{AB}(\mu,Q)\, \tilde J_{AB}(\mu,\nu)\, \tilde S_{AB}(\mu,\nu)\,,
\end{align}
where $\nu$ is the Laplace conjugate of $ d ^2$.  Over-tildes denote the Laplace transform, where, for the example of the jet function, is defined as
\begin{align}
\tilde J_{AB}(\mu,\nu) = \int d (d ^2)\, J_{AB}(\mu, d ^2)\,e^{-\nu d ^2}\,.
\end{align}

Renormalization in Laplace space is accomplished by linear, homogeneous differential equations.  The hard, jet, and soft functions satisfy the equations
\begin{align}
\mu\frac{\partial}{\partial \mu}H_{AB}(\mu,Q) &= \gamma_H(\mu,Q)\,H_{AB}(\mu,Q)\,,\\
\mu\frac{\partial}{\partial \mu}\tilde J_{AB}(\mu,\nu) &= \gamma_J(\mu,\nu)\,\tilde J_{AB}(\mu,\nu)\,,\nonumber\\
\mu\frac{\partial}{\partial \mu}\tilde S_{AB}(\mu,\nu) &= \gamma_S(\mu,\nu)\,\tilde S_{AB}(\mu,\nu)\,.\nonumber
\end{align}
The $\gamma_F$ functions are the anomalous dimensions for function $F$, which can be calculated order-by-order in the strong coupling.  Consistency of the factorization demands that the distribution is independent of the renormalization scale $\mu$, which is equivalent to the sum of anomalous dimensions vanishing:
\begin{align}
\gamma_H(\mu,Q)+\gamma_J(\mu,\nu)+\gamma_S(\mu,\nu) =0\,.
\end{align}

In the results we will show in future sections, we will perform the resummation to next-to-leading logarithmic accuracy.  To do this, requires the anomalous dimensions to one-loop accuracy.  These anomalous dimensions turn out to be especially simple and well-known, due to the limited possible correlations that can appear at one-loop order.  To demonstrate this, let's focus on the jet function and expand it to one-loop order:
\begin{align}
&J_{AB}( d ^2)\\
&=\int d\Pi_A\, d\Pi_B\, J(\Pi_A)\, J(\Pi_B)\,\delta\left(
 d ^2 - \text{SEMD}(\Pi_A,\Pi_B)
\right)\nonumber\\
&=\delta( d ^2)+2\int ds\, J^{(1)}(s)\,\delta( d ^2-s)+\cdots\,,\nonumber
\end{align}
where the superscript $(1)$ denotes the one-loop jet function or splitting function.  Through one-loop order, no correlations can appear because only one of the jets is allowed to have non-trivial structure.  Additionally, through this order, the SEMD is therefore equivalent to the jet mass.  Therefore, the one-loop renormalization of the jet function is completely governed by the one-loop anomalous dimension for the jet mass, which is extremely well-studied \cite{Catani:1992ua}.  Note that because this is the two-point correlation, the one-loop contribution is twice as large as one would find for the jet mass.

Therefore, the renormalization group equation for the jet function in Laplace space with one-loop running is
\begin{align}
\mu\frac{\partial}{\partial \mu}\tilde J_{AB}(\mu,\nu) &=2 \gamma_{J}^{(1)}(\mu,\nu)\,\tilde J_{AB}(\mu,\nu)\,.
\end{align}
The one-loop anomalous dimension for the jet mass is \cite{Bauer:2003pi,Bosch:2004th,Bauer:2001rh,Fleming:2003gt,Bauer:2006qp}
\begin{align}
\gamma_{J,q}^{(1)}(\mu^2,\nu) &=\frac{2\alpha_sC_F}{\pi}\log(\mu^2\nu e^{\gamma_E})+\frac{3C_F}{2}\frac{\alpha_s}{\pi}\,,\\
\gamma_{J,g}^{(1)}(\mu^2,\nu) &= \frac{2\alpha_sC_A}{\pi}\log(\mu^2\nu e^{\gamma_E})+\frac{\beta_0}{2}\frac{\alpha_s}{\pi}\,,
\end{align}
for quark and gluon jets, respectively.  $\gamma_E$ is the Euler-Mascheroni constant, $C_F=4/3$ and $C_A=3$ are the fundamental and adjoint quadratic Casimirs of SU(3) color, and $\beta_0$ is the coefficient of the one-loop $\beta$-function of QCD:
\begin{align}
\beta_0 = \frac{11}{3}C_A-\frac{2}{3}n_f\,,
\end{align}
where $n_f$ is the number of active quark flavors.

Complete results for accomplishing resummation at NLL accuracy with the approximations used in this paper are presented in App.~\ref{app:nllresum}.  One thing we note here is that to this order, the anomalous dimension of the SEMD, or almost any event metric, is the sum of anomalous dimensions of the jet mass on the two jets individually.  This simple sum form of the anomalous dimension was already noted at double logarithmic accuracy in Refs.~\cite{Komiske:2022vxg,Larkoski:2023qnv}, and here we see it holds more generally through NLL.  Beyond NLL, however, non-trivial correlations between the jets can and will generally be present.  In App.~\ref{app:twoloop}, we perform the jet function analysis through two-loop order, and show explicitly that the sum of jet mass anomalous dimensions is broken by correlations.  We present the calculation of the complete two-loop anomalous dimension, necessary for resummation through next-to-next-to-leading logarithm there, as well.  However, for all plots and numerical comparisons we will present in future sections, we will restrict our attention and predictions to NLL accuracy.

For higher-point correlations, these fundamental results for NLL resummation hold; one just has to carefully keep track of bookkeeping the explosion of indices.  Considering the three-point correlator in particular, the same exercise results in the following one-loop renormalization group evolution for the jet function in Laplace space:
\begin{align}\label{eq:3ptresum}
&\mu\frac{\partial}{\partial \mu}\tilde J_{ABC}(\mu,\nu_{AB},\nu_{AC},\nu_{BC}) \\
&=\left[
\gamma_J^{(1)}(\mu,\nu_{AB}+\nu_{AC})+\gamma_J^{(1)}(\mu,\nu_{AB}+\nu_{BC})\right.\nonumber\\
&\left.\hspace{1cm}+\gamma_J^{(1)}(\mu,\nu_{AC}+\nu_{BC})
\right] \tilde J_{ABC}(\mu,\nu_{AB},\nu_{AC},\nu_{BC})\nonumber\,.
\end{align}
The anomalous dimensions are still the same jet mass anomalous dimensions, but their argument is now a sum of a pair of Laplace conjugates to the corresponding SEMD values.  One-loop renormalization of general $n$-point correlations will naturally have a total of $n$ anomalous dimensions in the sum and in each, $n-1$ Laplace conjugates summed in the arguments.

\section{Application: Calculation of Dataspace Non-Gaussianities}\label{sec:nongauss}

With this resummation formalism for the event correlators established, we now move to applying it in an explicit example.  In particular, we will study and calculate the dataspace non-Gaussianity $\eta_\text{n-G}( d )$, introduced in Ref.~\cite{Larkoski:2025idq}.  This quantity is a measure of the extent to which bin-by-bin fluctuations of the two-point event correlator are non-Gaussian.  If $\eta_\text{n-G}( d ) = 0$, then fluctuations in the bin about distance $d$ are Gaussian, and further, the dataspace manifold exhibits a translational symmetry at that scale.

The event simulation results of Ref.~\cite{Larkoski:2025idq} also suggested that the non-Gaussianity $\eta_\text{n-G}( d )$ is sensitive to the transition from quarks and gluons as the degrees of freedom of QCD at large energy distances $d$, to hadrons, the degrees of freedom at small distances $d$.  Specifically, it was observed in simulation that around the scale of the parton-to-hadron transition $d_\text{non-pert}$, the non-Gaussianity exhibits a universality in that its value is independent of the overall jet or event energy.  That is, $\eta_\text{n-G}( d_\text{non-pert} )$ seemed to take a value intrinsic to the dynamics of confinement of QCD, independent of the high energy scale at which the process of interest occurs.

With a factorization theorem and framework for systematically-improvable calculations of the event correlators, we can validate this simulated empirical observation with a concrete prediction.  For the purposes of this paper, the dataspace non-Gaussianity $\eta_\text{n-G}( d )$ at distance $d$ is defined as
\begin{widetext}
\begin{align}
&\eta_\text{n-G}( d ) = \frac{\int d\Pi_A\, d\Pi_B\, d\Pi_C\, p(\Pi_A)\,p(\Pi_B)\,p(\Pi_C)\,\delta\left( d -d(\Pi_A,\Pi_B)\right)\delta\left( d -d(\Pi_A,\Pi_C)\right)}{\int d\Pi_A\, d\Pi_B\, p(\Pi_A)\,p(\Pi_B)\,\delta\left( d -d(\Pi_A,\Pi_B)\right)} \\
&\hspace{7cm}
- \int d\Pi_A\, d\Pi_B\, p(\Pi_A)\,p(\Pi_B)\,\delta\left( d -d(\Pi_A,\Pi_B)\right)\nonumber\,,
\end{align}
\end{widetext}
where $d(\Pi,\Pi')$ is the dataspace metric, evaluated between two points on phase space, $\Pi,\Pi'$, and $p(\Pi)$ is the distribution of events on phase space.\footnote{This differs slightly from the definition introduced in Ref.~\cite{Larkoski:2025idq}.  In that reference, the bin width $\delta d$ and number of events on the dataspace $N$ were also included as overall factors.  We omit these multiplicative factors as they are specific to the statistical analysis at hand, and not the underlying physics.}  If the dataspace exhibits a translation symmetry at the scale $d$, then the distribution of metric distances about any pair of points is identical.  That is, if there is a translation symmetry, we can freely set one point to be the (arbitrary) origin $\vec 0$, and so
\begin{align}
&\int d\Pi_A\, d\Pi_B\, p(\Pi_A)\,p(\Pi_B)\,\delta\left( d -d(\Pi_A,\Pi_B)\right)\\
& \hspace{1.5cm}\xrightarrow{\text{symmetry}} \int d\Pi\,p(\Pi)\,\delta\left( d -d(\vec 0,\Pi)\right)\,.\nonumber
\end{align}
In such a case, the non-Gaussianity vanishes.

To calculate the non-Gaussianity $\eta_\text{n-G}( d )$ within our factorization theorem, we note that it consists of the two- and three-point event correlators.  For the three-point event correlator, the non-Gaussianity needs the correlation between a point and two other points, each of which are the same distance $d$ from the initial point.  This can be calculated from the factorization theorem for the distribution of two pairwise distances, $d_{AB}$ and $d_{AC}$, and then setting $d_{AB}=d_{AC} = d$.  Within our factorization theorem framework, the relevant distribution is
\begin{align}
&\hspace{-0.2cm}p( d _{AB}^2, d _{AC}^2)=H_{ABC}\int d d _{AB,J}^2\,d d _{AC,J}^2\,d d _{AB,S}^2\,d d _{AC,S}^2\,\nonumber\\
&\hspace{0cm}\times J_{ABC}( d _{AB,J}^2, d _{AC,J}^2)\, S_{ABC}( d _{AB,S}^2, d _{AC,S}^2)\\
&\hspace{0cm}\times \delta( d_{AB}^2- d_{AB,J}^2- d_{AB,S}^2) \delta( d_{AC}^2- d_{AC,J}^2- d_{AC,S}^2)
\nonumber\,,
\end{align}
where $H_{ABC}$, $J_{ABC}( d _{AB,J}^2, d _{AC,J}^2)$, and $S_{ABC}( d _{AB,S}^2, d _{AC,S}^2)$ are the hard, jet, and soft functions for this process, as defined in the previous section.

Transforming this distribution to Laplace space, it takes the product form
\begin{align}
\tilde p(\nu_{AB},\nu_{AC})&=H_{ABC}(\mu^2,Q^2)\\
&\hspace{-1cm}\times \tilde J_{ABC}(\mu^2,\nu_{AB},\nu_{AC})\, \tilde S_{ABC}(\mu^2,\nu_{AB},\nu_{AC})\nonumber\,,
\end{align}
where over-tildes represent Laplace transformed functions, and $\nu_{AB}$ and $\nu_{AC}$ are the Laplace conjugates of squared distances $d_{AB}^2$ and $d_{AC}^2$, respectively.  Integration over an observable is equivalent to setting its Laplace conjugate to 0.  Therefore, the renormalization group evolution equation for the jet function, for example, is simply determined from Eq.~\ref{eq:3ptresum} by setting $\nu_{BC} = 0$.  That is,
\begin{align}
&\mu^2\frac{\partial}{\partial \mu^2}\tilde J_{ABC}(\mu^2,\nu_{AB},\nu_{AC}) \\
&\hspace{1cm}=\left[
\gamma_J^{(1)}(\mu^2,\nu_{AB}+\nu_{AC})+\gamma_J^{(1)}(\mu^2,\nu_{AB})\right.\nonumber\\
&\left.\hspace{2cm}+\gamma_J^{(1)}(\mu^2,\nu_{AC})
\right] \tilde J_{ABC}(\mu^2,\nu_{AB},\nu_{AC})\nonumber\,,
\end{align}
where $\gamma_J^{(1)}(\mu^2,\nu)$ is the one-loop anomalous dimension of the squared jet mass in Laplace space.

Explicit calculations of all the necessary ingredients for resummation to next-to-leading logarithmic (NLL) accuracy of the non-Gaussianity are presented in App.~\ref{app:nllresum}. In the following section, we present results and plots of these predictions.  One particularly interesting and subtle aspect of the non-Gaussianity, or specifically, the particular three-point correlator present within it, is that it is not IRC safe.  The measurement of two distances on dataspace is IRC safe, given an IRC safe metric like the SEMD.  As such, non-zero values of both distances regulate infrared and collinear divergences in a fixed-order calculation.  However, in the non-Gaussianity, we take those two distances to be equal.  That is, at lowest order in perturbation theory, the distribution $p(d_{AB}^2,d_{AC}^2)$ takes the form
\begin{align}
p(d_{AB}^2,d_{AC}^2) = \delta\left(d_{AB}^2\right)\delta\left(d_{AC}^2\right)+{\cal O}(\alpha_s)\,,
\end{align}
which is unit normalized when integrated over $d_{AB}^2,d_{AC}^2$. Setting the two distances equal produces a non-integrable distribution in the common distance $d^2$:
\begin{align}
\left.p(d_{AB}^2,d_{AC}^2)\right|_{d_{AB}^2=d_{AC}^2=d^2} = \delta\left(d^2\right)\delta\left(d^2\right)+{\cal O}(\alpha_s)\,.
\end{align}
Nevertheless, we can resum the distribution $p(d_{AB}^2,d_{AC}^2)$ in both distances, and then set them equal.  Performing the resummation first regulates and in fact exponentially suppresses the singular phase space region that is problematic at fixed order.  As such, the non-Gaussianity is an example of a Sudakov safe observable \cite{Larkoski:2013paa,Larkoski:2015lea,Larkoski:2014wba}, which cannot be calculated in fixed-order perturbation theory, but can be calculated in resummed perturbation theory.  This feature of the non-Gaussianity may have additional consequences for the dataspace manifold, but we leave a further study in this direction to future work.

\subsection{Two-Point Event Correlator}

\begin{figure}[t!]
\begin{center}
\includegraphics[width=0.45\textwidth]{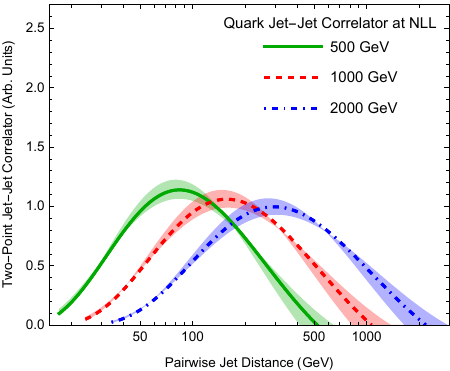}
\caption{\label{fig:jjq_nll}
Plots of the two-point correlator measured on pairs of quark jets, resummed at NLL accuracy, as a function of pairwise jet distance, $d$.  The distribution is plotted logarithmically in distance $d$, $d\, p(d)$, for jets with three different energies: 500 GeV (solid green), 1000 GeV (dashed red), 2000 GeV (dot-dashed blue), and the shading is representative of residual scale dependence from neglecting higher-order contributions.
}
\end{center}
\end{figure}

\begin{figure}[t!]
\begin{center}
\includegraphics[width=0.45\textwidth]{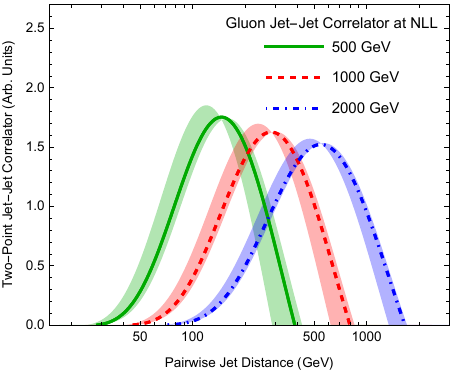}
\caption{\label{fig:jjg_nll}
Plots of the two-point correlator measured on pairs of gluon jets, resummed at NLL accuracy, as a function of pairwise jet distance, $d$.  The distribution is plotted logarithmically in distance $d$, $d\, p(d)$, for jets with three different energies: 500 GeV (solid green), 1000 GeV (dashed red), 2000 GeV (dot-dashed blue), and the shading is representative of residual scale dependence from neglecting higher-order contributions.
}
\end{center}
\end{figure}

Before presenting results for the non-Gaussianity, we first show distributions from calculations of the two-point correlator measured on pairs of jets.  These are displayed in Figs.~\ref{fig:jjq_nll} and \ref{fig:jjg_nll}, for the two-point jet correlator measured on quark and gluon jet samples, respectively.  In these and all explicit results presented in this paper, we work in the collinear limit in which the description of quark and gluon jets is process independent.  In each of these plots, we show the jet-jet correlator at three different jet energies, 500, 1000, and 2000 GeV, to observe scaling violations.  Each distribution is normalized to isolate shape variation.  Within the context of the factorization theorem, we can vary the scales about their canonical values, to illustrate residual scale dependence from truncating the calculation to finite logarithmic accuracy.  We do not attempt a general scale variation or perturbative uncertainty analysis here, but merely vary the scale of the soft function by a factor of two up and down about its canonical value, illustrated by the shaded bands.

As explicitly demonstrated in earlier sections, the two-point jet-jet correlator through the SEMD is simply the sum of the squared masses of the jets.  As such, the shape and scale sensitivity of these distributions is familiar from the numerous related studies of the jet mass or thrust observables.  However, what is very different between the familiar jet mass and the two-point jet-jet correlator is that the calculation of the latter requires two probability distributions of the jets on phase space.  Each distribution on phase space provides exponential Sudakov suppression, and so the two-point jet-jet correlator is Sudakov suppressed squared.  This observation has been observed before at double logarithmic accuracy \cite{Komiske:2022vxg,Larkoski:2023qnv}, but its consequence for systematically-improvable predictions and at higher accuracy is rather profound.

In particular, this extra Sudakov suppression means that a perturbative analysis is valid over nearly the entire domain over which the two-point jet-jet correlator has support.  Non-perturbative physics dominates the SEMD when the value of the SEMD or squared mass from the soft function becomes non-perturbative.  This contribution to the squared mass from a soft emission near the boundary of the jet is:
\begin{align}
d_s^2 \sim E_sER^2\,,
\end{align}
where $E_s$ is the soft emission's energy, $E$ is the total jet energy, and $R$ is the jet radius.  The soft emission becomes non-perturbative when its transverse momentum from the jet axis is comparable to the QCD scale, $\Lambda_\text{QCD}$:
\begin{align}
\Lambda_\text{QCD}\sim E_s R\,.
\end{align}
Combining these expressions, the value of the inter-jet distance at which non-perturbative physics dominates is
\begin{align}
d_\text{np} \sim\sqrt{\Lambda_\text{QCD}ER}\,.
\end{align}
All curves in both Figs.~\ref{fig:jjq_nll} and \ref{fig:jjg_nll} are plotted down to this minimal perturbative scale, with $\Lambda_\text{QCD}\sim 1$, and taking the jet radius to be order-1, $R\sim 1$.

As demonstrated in the plots, this scale is well below the peaks of the distributions.  Comparison of theoretical predictions with experimental data therefore requires minimal non-perturbative input and the fact that the entire peak region of the distribution is perturbative may have significant consequences for precision extractions of the strong coupling, $\alpha_s$.  Nevertheless, a sufficiently focused study of the small distance regime would require something like a shape function \cite{Korchemsky:1999kt,Korchemsky:2000kp}, which would be necessary to robustly demonstrate scaling relations that relate finite dataset size to interjet distances on the dataspace manifold \cite{Batson:2023ohn,Larkoski:2025clo}.

\subsection{Non-Guassianities Near the QCD Scale}

We continue and present calculations of the non-Gaussianity measure $\eta_\text{n-G}(d)$ on these same jet samples.  The non-Gaussianity observable $\eta_\text{n-G}(d)$ contains the three-point jet-jet-jet correlator and calculation of this object in resummed perturbation theory is significantly more subtle than that of the two-point correlator.  Specifically, the two-point correlator can be expressed exactly in closed form in terms of exponential functions, Gamma functions, and their derivatives, as shown in App.~\ref{app:nllresum}.  However, for the three-point correlator, while resummation can be easily accomplished in Laplace space through solving the renormalization group equations, performing the inverse Laplace transform back to observable space is extremely challenging.  No known exact expression for the inverse Laplace transform necessary to resum the three-point correlator is known, so we rely on approximations.  Specifically, we use Post's approximation for the inverse Laplace transform \cite{post1930generalized,widder1934inversion}, with all details regarding this approximation and its practical implementation presented in App.~\ref{app:nllresum}.

\begin{figure}[t!]
\begin{center}
\includegraphics[width=0.45\textwidth]{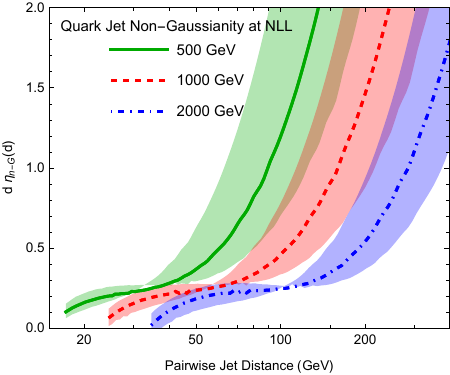}
\caption{\label{fig:ngq_nll}
Plot of the non-Gaussianity measure logarithmic in SEMD distance, $d\, \eta_\text{n-G}(d)$ as measured on quark jets resummed to NLL accuracy.  Three different jet energies are presented: 500 GeV (solid green), 1000 GeV (dashed red), and 2000 GeV (dot-dashed blue), and shaded bands are representative of residual scale dependence from truncating the resummed perturbation series.
}
\end{center}
\end{figure}

\begin{figure}[t!]
\begin{center}
\includegraphics[width=0.45\textwidth]{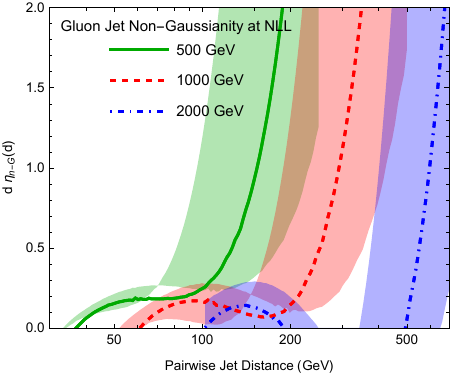}
\caption{\label{fig:ngg_nll}
Plot of the non-Gaussianity measure logarithmic in SEMD distance, $d\, \eta_\text{n-G}(d)$ as measured on gluon jets resummed to NLL accuracy.  Three different jet energies are presented: 500 GeV (solid green), 1000 GeV (dashed red), and 2000 GeV (dot-dashed blue), and shaded bands are representative of residual scale dependence from truncating the resummed perturbation series.
}
\end{center}
\end{figure}

Results of the calculation of the non-Gaussianity $\eta_\text{n-G}(d)$ are presented in Figs.~\ref{fig:ngq_nll} and \ref{fig:ngg_nll}, for quark and gluon jets, respectively, at the same three jet energies studied in the previous section: 500, 1000, and 2000 GeV.  As with the plots of the two-point correlator, we also vary the soft scale up and down by a factor of 2 as representative of residual scale dependence from truncation to finite logarithmic perturbative accuracy.  This scale dependence is illustrated by the shaded bands on these plots.

Especially in Fig.~\ref{fig:ngq_nll} for the non-Gaussianity measured on quark jets, there is a rather remarkable universality of a ledge or shelf, broadly independent of jet energy.  Around the point where the slope in the non-Gaussianity vanishes, it takes the value of about 0.2 on each sample, suggestive of some underlying physical property of the jets responsible for this structure.  This universal ledge of the non-Gaussianity was also observed in simulated jets in Ref.~\cite{Larkoski:2025idq}, where it was conjectured that it was related to the parton-to-hadron transition.  However, here we observe that this is actually a purely perturbative phenomena, as no non-perturbative information has been included in these calculations (other than terminating at sufficiently low scales where the coupling $\alpha_s$ would diverge).  Additionally, this ledge structure is truly due to the detailed interplay of the two- and three-point correlators.  Comparing the location of the ledges to the two-point correlators in Fig.~\ref{fig:jjq_nll}, it appears that the ledges may correspond to inflection points there.

To emphasize this point, we recall the simple non-perturbative scaling analysis from the previous section.  Non-perturbative physics is expected to dominate at and below a scale $d_\text{np}\sim \sqrt{\Lambda_\text{QCD}E R}$.  For the jet energies we consider here and assuming that the jet radius $R=1$, the non-perturbative distances are approximately $d_\text{np}\sim 22$, $32$, and $45$ GeV, for jet energies of 500, 1000, and 2000 GeV, respectively.  By contrast, the ledges in the non-Gaussianity distributions appear to lie at about 30, 50, and 80 GeV, respectively, seemingly comfortably above the corresponding non-perturbative scales.  However, the distance scale of the ledges is clearly not parametrically larger than the non-perturbative scale, which is why Ref.~\cite{Larkoski:2025idq} conjectured a deeper relationship.

While our resummed perturbative calculations are sensible and likely dominate in the region of the ledges, there still may be significant, though formally subdominant, non-perturbative effects contributing.  For example, in the case of the jet mass, the effect of non-perturbative physics in the region where a perturbative description dominates is well-known to be a lateral displacement of the distribution \cite{Dokshitzer:1995zt,Lee:2006fn}.  A similar non-perturbative translation of the non-Gaussianity may be important for the ledges, and could shift its location to higher, or lower, distances.  More work is needed to conclusively demonstrate that the structure in the non-Gaussianity is dominantly perturbative, correlated with the second derivatives of the two-point correlator, or to establish the extent to which non-perturbative physics affect it, however.

Turning to the non-Gaussianities on gluon jet samples, Fig.~\ref{fig:ngg_nll}, this interpretation is much more subtle to make.  The 500 GeV gluon jet sample exhibits the ledge structure, but at higher energies, the ledge is significantly more curvy, or the non-Gaussianity even takes (unphysical) negative values.  These features of the gluon non-Gaussianities can be traced to the limitations of approximation used to evaluate the inverse Laplace transform of the three-point correlator.  The approximations we use are most accurate at low scales, and deviate significantly at high scales, where logarithms are small.  Sensitivity to higher scales and therefore smaller logarithms is more pronounced in the gluon jet samples because of the larger color factor of gluon jets, as compared to quark jets.  The larger color factor for gluons produces stronger Sudakov suppression, pushing the peak of the two-point correlator on gluon jets to higher scales than that for quark jets. In App.~\ref{app:3pt}, we demonstrate explicitly that this is indeed due to the finite-order truncation of Post's approximation for the inverse Laplace transformation.  This motivates further work on solving this limitation once and for all.

%\section{Comparison with Event Simulation}

\section{Conclusions}\label{sec:concs}

In this paper, we have presented the first systematically-improvable factorization framework for calculation of the geometry of collider physics event dataspace.  Asking geometric questions about the dataspace manifold requires an IRC safe metric, and the features of the SEMD, its invariance to isometries, its closed-form expression, and its fast evaluation, make it the currently-preferred choice of metric.  Factorization of soft and collinear contributions to $n$-point event corrrelators enables a factorization theorem in the limit of small SEMD metric distance, and resummation of logarithms can be accomplished through renormalization group evolution.  We presented results for next-to-leading logarithmic resummation of the two- and three-point event correlators, and for the latter, focused on the measure of non-Gaussian fluctuations that can be defined from it.

Significant improvements to the calculations presented here can be accomplished in the near future.  Because of the close relationship between the $p=2$ SEMD and the jet mass, all ingredients for resummation to next-to-next-to-leading logarithm of the two-point event correlator are known.  Additionally, the extra Sudakov suppression renders its distribution perturbative over nearly the entire domain of support, and so could produce high-precision extraction of the strong coupling, $\alpha_s$.  However, incorporating high-precision fixed-order contributions into the event correlator prediction may be a significant challenge.  Standard IRC safe observables are purely a function of the momenta of particles in a single event, and so techniques for real-virtual divergence cancellation, like subtraction methods \cite{Catani:1996vz,Frixione:1995ms}, are applied event-by-event.  The meta-event structure of event dataspace correlators would require careful reorganization of divergences and consideration of their correlation across distinct events.

In fact, the dataspace metric and the minimal smearing distance analysis may provide a new approach to fixed-order divergence generation and cancellation.  Virtual and real contributions to a process live on different particle multiplicity phase spaces.  Except exactly at the singular point, these contributions will have a non-zero metric distance from one another and that distance is itself a measure of the precision to which the divergences are canceled.  In the context of more familiar fixed-order methods, this metric distance between real and virtual contributions would be like the bin width in some distribution of a specific IRC safe observable.  However, because of the identity of indiscernibles property, the metric distance would regulate divergences for any possible IRC safe observable.  In this way, these leading-power, factorized, and resummed results for the two-point event correlator may be particularly useful for extending methods like that introduced in Refs.~\cite{Catani:2007vq,Gao:2012ja,Gao:2014nva,Boughezal:2015dva,Boughezal:2015aha,Gaunt:2015pea,Moult:2016fqy} for high precision fixed-order predictions.

Continuing predictions of three- and higher-point event correlators, the resummation structure in Laplace space is similar, but the inverse Laplace transform back to observable space is highly non-trivial.  Even for the three-point correlator or the non-Gaussianity measure, standard approximations to the inverse Laplace transformation can be insufficient to render predictions non-negative over the entire phase space.  So, extending predictions beyond the two-point event correlator will require better approximations for inverse Laplace transformations, or introducing approaches to resummation that side-step transformation to a conjugate space.  Further advancements of techniques for precision calculations of the distribution of collider events on its manifold will shine a light on new corners of this vast space.

\acknowledgements

I thank Rikab Gambhir, Rishabh Jain, Matt LeBlanc, Jesse Thaler, and Grant Whitman for comments.

\appendix

\section{Anomalous Dimensions and NLL Resummation}\label{app:nllresum}

In this appendix, we present all results necessary for resummation of the $n$-point event correlations with the SEMD through next-to-leading logarithmic accuracy.

\subsection{One-Loop Anomalous Dimensions}

In Sec.~\ref{sec:oneloopresum}, we presented the one-loop anomalous dimensions of the jet function of the invariant mass.  For the factorization theorem, we need also the corresponding anomalous dimensions of the soft function and the hard function.  For simplicity, we work in the boosted limit, in which the jet of interest recoils against the entire rest of the event, which is assumed to be collimated in the direction opposite to the jet.  As such, the soft function on which we measure the squared jet mass can be calculated from the soft limit of the jet function, along with the constraint that the soft emission is contained in the jet of radius $R$.  This one-loop soft function is
\begin{align}
S^{(1)}(s) &= \frac{\alpha_sC_J}{\pi}\frac{\mu^{2\epsilon}}{\Gamma(1-\epsilon)}\frac{1}{s^{1+\epsilon}}\int \frac{dz}{z^{1+\epsilon}} \,\Theta(zE^2R^2-s)\nonumber\\
&=\frac{\alpha_sC_J}{\pi}\frac{\mu^{2\epsilon}}{\Gamma(1-\epsilon)}\frac{1}{\epsilon}\frac{(E^2R^2)^\epsilon}{s^{1+2\epsilon}}\,,
\end{align}
where the spacetime dimension is $D=4-2\epsilon$.  Here, $C_J$ is the color Casimir of the jet of interest.  We can take the Laplace transform of this expression to then identify the one-loop anomalous dimension, where
\begin{align}
\gamma_S^{(1)}(\mu,\nu) = -\frac{\alpha_sC_J}{\pi}\log\left(\mu^2\nu^2E^2R^2e^{2\gamma_E}\right)\,.
\end{align}

Then, given the results of the jet and soft function anomalous dimensions for the squared jet mass, we simply define the hard function's anomalous dimension to be that which renders the renormalization group evolution consistent.  We thus have, for quark and gluon jets, respectively, 
\begin{align}
\gamma_{H,q}^{(1)}(\mu,ER) &= -\gamma_{J,q}^{(1)}(\mu,\nu)-\gamma_{S,q}^{(1)}(\mu,\nu)\\
&=-\frac{\alpha_sC_F}{\pi}\log\frac{\mu^2}{E^2R^2}-\frac{3C_F}{2}\frac{\alpha_s}{\pi}\,,\nonumber\\
\gamma_{H,g}^{(1)}(\mu,ER) &= -\gamma_{J,g}^{(1)}(\mu,\nu)-\gamma_{S,g}^{(1)}(\mu,\nu)\\
&=-\frac{\alpha_sC_A}{\pi}\log\frac{\mu^2}{E^2R^2}-\frac{\beta_0}{2}\frac{\alpha_s}{\pi}\,.\nonumber
\end{align}

\subsection{Ingredients for Resummation}

To resum through NLL accuracy, there are a few universal quantities that are needed.  We need the cusp anomalous dimension through two-loop order.  We define the perturbative expansion of the cusp anomalous dimension as
\begin{align}
\Gamma_\text{cusp}(\alpha_s) = \sum_{i=0}^\infty \Gamma_i\left(
\frac{\alpha_s}{4\pi}
\right)^{i+1}\,,
\end{align}
where the $\Gamma_n$ are the series coefficients.  Through two-loops, this is \cite{Korchemsky:1987wg}
\begin{align}
\Gamma_0 &= 4\,,\\
\Gamma_1 &= 4C_A\left(
\frac{67}{9}-\frac{\pi^2}{3}
\right)-\frac{40}{9}n_f\,.
\end{align}
We also need the QCD $\beta$-function through two-loops.  The $\beta$-function and its perturbative expansion is
\begin{align}
\beta(\alpha_s) =\mu\frac{\partial \alpha_s}{\partial \mu} = -2\alpha_s\sum_{i=0}^\infty \beta_i\left(
\frac{\alpha_s}{4\pi}
\right)^{i+1}\,,
\end{align}
and the coefficients through two-loops are \cite{Tarasov:1980au}
\begin{align}
\beta_0&=\frac{11}{3}C_A-\frac{2}{3}n_f\,,\\
\beta_1&=\frac{34}{3}C_A^2-2n_f\left(
C_F+\frac{5}{3}C_A
\right)\,.
\end{align}

Then, we can explicitly solve the renormalization group evolution equations given the anomalous dimension.  In general, as relevant for $n$-point event correlation, the renormalization group evolution for a function $F_n(\mu)$ takes the general form
\begin{align}
\mu\frac{\partial}{\partial \mu}F_n(\mu) = n\gamma_F(\mu)\, F_n(\mu)\,.
\end{align}
Here, the anomalous dimension $\gamma_F(\mu)$ in the context at hand, is the appropriate anomalous dimension of the observable measured on a single event.  It takes the general form
\begin{align}
\gamma_F(\mu) = d_F \Gamma_\text{cusp}(\alpha_s)\log\frac{\mu^2}{\mu_1^2}+\gamma_F(\alpha_s)\,,
\end{align}
where $d_F$ is a constant, independent of $\alpha_s$, factor specific to the particular function $F$ under consideration and the types of jets in the process.  $\gamma_F(\alpha_s)$ is the non-cusp anomalous dimension, which has the perturbative expansion
\begin{align}
\gamma_F(\alpha_s) = \sum_{i=0}^\infty\gamma_i\left(
\frac{\alpha_s}{4\pi}
\right)^{i+1}\,.
\end{align}
The scale $\mu_1$ is the function-specific scale whose dependence is resummed to all orders in perturbation theory by the renormalization group evolution.

In Laplace space, where the factorization theorem expresses the cross section as a product of functions, the solution to the renormalization group evolution is, e.g., Refs.~\cite{Korchemsky:1993uz,Balzereit:1998yf,Neubert:2005nt,Becher:2006mr,Fleming:2007xt,Ellis:2010rwa,Frye:2016aiz},
\begin{align}
F_n(\mu) = e^{nK_F(\mu,\mu_0)}F_n(\mu_0)\left(
\frac{\mu_0^2}{\mu_1^2}
\right)^{n\omega_F(\mu,\mu_0)}\,.
\end{align}
In this function, $\mu_0$ is the natural scale of the function $F$, which is an order-one factor from the scale $\mu_1$; that is, no logarithms between scales $\mu_0$ and $\mu_1$ are resummed.  $F_n(\mu_0)$ is the boundary value of the function $F_n$ at the natural scale.  Through next-to-leading logarithmic accuracy, the functions $K(\mu,\mu_0)$ and $\omega(\mu,\mu_0)$ are
\begin{widetext}
\begin{align}
K_F(\mu,\mu_0) &= d_F\frac{\Gamma_0}{2\beta_0^2}\left[
\frac{4\pi}{\alpha_s(\mu_0)}\left(
\log r+\frac{1}{r}-1
\right)+\left(
\frac{\Gamma_1}{\Gamma_0} - \frac{\beta_1}{\beta_0}
\right)(r-1-\log r)-\frac{\beta_1}{2\beta_0}\log^2 r
\right]-\frac{\gamma_0}{2\beta_0}\log r\,,\\
\omega_F(\mu,\mu_0)&=-d_F\frac{\Gamma_0}{2\beta_0}\left[
\log r+\frac{\alpha_s(\mu_0)}{4\pi}\left(
\frac{\Gamma_1}{\Gamma_0}-\frac{\beta_1}{\beta_0}
\right)(r-1)
\right]\,,
\end{align}
\end{widetext}
where $r$ is the ratio of the strong coupling evaluated at the two scales
\begin{align}
r = \frac{\alpha_s(\mu)}{\alpha_s(\mu_0)}\,.
\end{align}

\subsection{Resummation of the Two-Point Event Correlator}

Applying these general results to the two-point event correlator distribution, its NLL resummed form in Laplace space is
\begin{widetext}
\begin{align}
\tilde p(\nu)&=e^{
2[K_H(\mu,\mu_H)+K_J(\mu,\mu_J)+K_S(\mu,\mu_S)]
}H_{AB}(\mu_H)J_{AB}(\mu_J)S_{AB}(\mu_S)\\
&\hspace{3cm}\times\left(
\frac{\mu_H^2}{E^2R^2}
\right)^{2\omega_H(\mu,\mu_H)}\left(
\mu_J^2\nu e^{\gamma_E}
\right)^{2\omega_J(\mu,\mu_J)}\left(
\mu_S^2\nu^2E^2R^2e^{2\gamma_E}
\right)^{2\omega_S(\mu,\mu_S)}
\nonumber\,,
\end{align}
\end{widetext}
where the natural scales of the hard, jet, and soft functions are denoted as $\mu_H$, $\mu_J$, and $\mu_S$, respectively.  To determine the distribution in original, ``real'' space, we note two things.  First, the explicit dependence on the Laplace conjugate $\nu$ is in a power-law form, and its inverse Laplace transform is
\begin{align}
{\cal L}^{-1}[\nu^q] = \frac{(d^2)^{-q-1}}{\Gamma(-q)}\,,
\end{align}
where $q$ is the appropriate power and $\Gamma(x)$ is the Euler Gamma function.  Additionally, there is implicit logarithmic dependence on the Laplace conjugate $\nu$ in the jet and soft functions.  However, we note that the derivative 
\begin{align}
\frac{\partial}{\partial q}\nu^q = \nu^q\log\nu\,,
\end{align}
commutes with the inverse Laplace transform.  So, we can replace all explicit low-scale logarithms with the appropriate derivatives and perform the inverse Laplace transform to produce:
\begin{widetext}
\begin{align}
p(d^2)&=\frac{1}{d^2}\,e^{
2[K_H(\mu,\mu_H)+K_J(\mu,\mu_J)+K_S(\mu,\mu_S)]
}H_{AB}(\mu_H)J_{AB}\left(L\to\frac{\partial_{\omega_J}}{2}\right)S_{AB}\left(L\to\frac{\partial_{\omega_S}}{2}\right)\\
&\hspace{1cm}\times\left(
\frac{\mu_H^2}{E^2R^2}
\right)^{2\omega_H(\mu,\mu_H)}\left(
\frac{\mu_J^2e^{\gamma_E}}{d^2} 
\right)^{2\omega_J(\mu,\mu_J)}\left(
\frac{\mu_S^2E^2R^2e^{2\gamma_E}}{d^4}
\right)^{2\omega_S(\mu,\mu_S)}\frac{1}{\Gamma\left(
-2\omega_J(\mu,\mu_J)-4\omega_S(\mu,\mu_S)
\right)}\,.
\nonumber
\end{align}
\end{widetext}
In the jet and soft functions, the notation
\begin{align}
L\to\frac{\partial_{\omega}}{2}\,,
\end{align}
means that we replace the logarithm $L$ by the corresponding derivative.

\subsection{Resummation of the Three-Point Event Correlator}\label{app:3pt}

Performing the analogous analysis for the three-point event correlator, we can express the Laplace transformed distribution as
\begin{widetext}
\begin{align}
\tilde p(\nu_{AB},\nu_{AC})&=e^{
3[K_H(\mu,\mu_H)+K_J(\mu,\mu_J)+K_S(\mu,\mu_S)]
}H_{ABC}(\mu_H)J_{ABC}(\mu_J)S_{ABC}(\mu_S)\\
&\hspace{3cm}\times\left(
\frac{\mu_H^2}{E^2R^2}
\right)^{3\omega_H(\mu,\mu_H)}\left(
\mu_J^2\left(\nu_{AB}\nu_{AC}(\nu_{AB}+\nu_{AC})\right)^{1/3} e^{\gamma_E}
\right)^{3\omega_J(\mu,\mu_J)}\nonumber\\
&\hspace{3cm}\times\left(
\mu_S^2\left(\nu_{AB}\nu_{AC}(\nu_{AB}+\nu_{AC})\right)^{2/3}E^2R^2e^{2\gamma_E}
\right)^{3\omega_S(\mu,\mu_S)}
\nonumber\,.
\end{align}
\end{widetext}
Here, recall that $\nu_{AB}$ and $\nu_{AC}$ are the Laplace conjugates of the squared pairwise distances $d_{AB}^2$ and $d_{AC}^2$.  As with the two-point energy correlator, we need to take the inverse Laplace transform in these two conjugate variables.  Then, for application to evaluation of the non-Gaussianities on dataspace, we set the distances $d_{AB}^2 = d_{AC}^2=d^2$.

The general function of the two Laplace conjugates that we need to invert takes the form
\begin{align}\label{eq:invlap3pt}
\left.{\cal L}_{\nu_1,\nu_2}^{-1}[\nu_1^q\nu_2^q(\nu_1+\nu_2)^q]\right|_{d_1^2=d_2^2=d^2}=(d^2)^{-2-3q}f(q)\,,
\end{align}
where the power to which $d^2$ is raised follows from dimensional analysis, and $f(q)$ is some function of the power $q$ to be determined.  To determine this function, note that the inverse Laplace transform over just $\nu_2$ is
\begin{align}
&\left.{\cal L}_{\nu_2}^{-1}[\nu_1^q\nu_2^q(\nu_1+\nu_2)^q]\right|_{d_2^2=1} \\
&\hspace{2cm}=\sqrt{\pi}\,\frac{e^{-\frac{\nu_1}{2}} \nu_1^{2q+1/2}I_{-q-1/2}\left(
\frac{\nu_1}{2}
\right)}{\Gamma(-q)}\,,\nonumber
\end{align}
where we can set $d_1^2=1$ to isolate the $q$ dependence, and $I_\alpha(z)$ is the modified Bessel function of the first kind.  To the best of our knowledge, there is no closed-form expression for the inverse Laplace transform of this expression.  However, in practice, we can use Post's inversion formula \cite{post1930generalized,widder1934inversion}.  If $G(\nu)$ is the Laplace transform of $g(t)$, then, the inverse Laplace transform can be expressed as
\begin{align}
g(t) = \lim_{k\to \infty}\frac{(-1)^k}{k!}\left(\frac{k}{t}\right)^{k+1}\left.\frac{d^k}{d\nu^k}G(\nu)\right|_{\nu = \frac{k}{t}}\,.
\end{align}

In practice, we cannot take an infinite number of derivatives, so we will take $k=25$.  We will use
\begin{align}
f(q)=\frac{(-1)^k}{k!}k^{k+1}\!\left.\frac{d^k}{d\nu^k}F(\nu)\right|_{\nu = k=25}\,,
\end{align}
in Eq.~\ref{eq:invlap3pt}, with
\begin{align}\label{eq:laplcefunc3pt}
F(\nu) = \sqrt{\pi}\,\frac{e^{-\frac{\nu}{2}} \nu^{2q+1/2}I_{-q-1/2}\left(
\frac{\nu}{2}
\right)}{\Gamma(-q)}\,.
\end{align}
Additionally, the asymptotic form of the Bessel function can be used to establish the inverse Laplace transform in the limit that $-q\to\infty$.  In this limit, we have
\begin{align}\label{eq:besselasymp}
\lim_{-q\to\infty}I_{-q-1/2}\left(
\frac{\nu}{2}
\right) = \frac{\left(\frac{\nu}{4}\right)^{-q-1/2}}{\Gamma\left(
\frac{1}{2}-q
\right)}\,.
\end{align}
The inverse Laplace transform in this limit is therefore
%\begin{align}\label{eq:invlaptransasymp}
%\lim_{-q\to\infty}f(q) = \frac{2^{2+3q}\sqrt{\pi}}{\Gamma\left(
%\frac{1}{2}-q
%\right)\Gamma(-q)^2}\,.
%\end{align}
\begin{align}\label{eq:invlaptransasymp}
\lim_{-q\to\infty}f(q) = \frac{2^{1+q}}{\Gamma(-2q)\Gamma(-q)}\,.
\end{align}

\begin{figure}[t!]
\begin{center}
\includegraphics[width=0.45\textwidth]{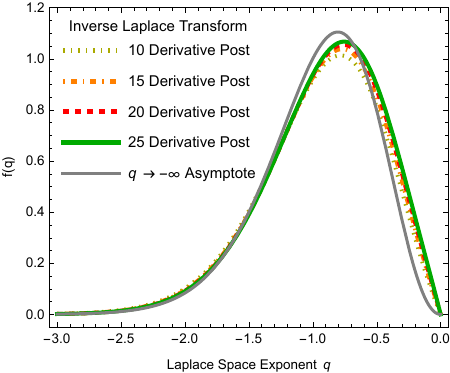}
\caption{\label{fig:post}
Plot of $f(q)$ as a function of exponent $q$ in the inverse Laplace transform of the resummed three-point event correlator.  Different approximations to $f(q)$ are plotted, as defined by the order $k$ of the derivative used in Post's formula: $k=10$ (dotted yellow), $k=15$ (dot-dashed orange), $k=20$ (dashed red), and $k=25$ (sold green).  Also plotted is the asymptotic form of $f(q)$, valid in the limit $-q\to\infty$ (solid gray).
}
\end{center}
\end{figure}

Good convergence of the functional form of the inverse Laplace transform is observed, as the number of derivatives $k$ in Post's formula increases.  We show this in Fig.~\ref{fig:post}, where we plot the functional form of $f(q)$ from Post's formula with $k=10,15,20,$ and 25 derivatives.  On this figure, we also plot the asymptotic inverse Laplace transform of Eq.~\ref{eq:invlaptransasymp}.  The Post formula approximations agreed well with the asymptotic expression by around $-q\simeq 1.5$ and beyond.

At the opposite limit, around $-q\to 0$, I am not aware of a robust approximation for the Bessel Function.  A simple visual inspection of the Post approximation suggests that the inverse Laplace transform $f(q)$ is approximately linear in this regime.  Further, the exponent $q$ has a Taylor series in the strong coupling $\alpha_s$ (through the $\omega$ functions), so this would correspondingly imply that the behavior of $f(q)$ about $q = 0$ could be calculated in fixed-order perturbation theory.  However, this would contradict our expectation that this observable is merely Sudakov safe, and not fully IRC safe, so there must be something we are missing.

\begin{figure}[t!]
\begin{center}
\includegraphics[width=0.45\textwidth]{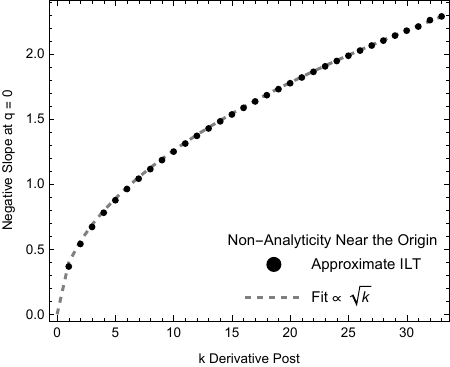}
\caption{\label{fig:derdiv}
Plot of the negative slope of the inverse Laplace transform (ILT) $f(q)$ at the origin, $q=0$, as a function of the derivative degree $k$ of Post's approximation.  The points are fit to a curve proportional to $\sqrt{k}$ (dashed gray).
}
\end{center}
\end{figure}

This can be observed through evaluation of the derivative of $f(q)$ at $q=0$, as a function of the Post approximation degree $k$.  This is plotted in Fig.~\ref{fig:derdiv}.  If the inverse Laplace transform $f(q)$ were truly linear in $q$ about the origin, then we would expect convergence of the derivative as $k$ increases.  By contrast, what is observed is that the slope at the origin continues to increase as $k$ increases, and to very good approximation, follows a fit proportional to $\sqrt{k}$.  This suggests that as $k\to\infty$ in the Post approximation, the derivative of $f(q)$ diverges at $q=0$, implying that it lacks a Taylor expansion about the origin.  This is then consistent with our expectation of Sudakov safety.  However, we would still desire an appropriate approximation for the Bessel function in this regime, so that the true behavior of the function $f(q)$ could be identified.  We look forward to this in future work.\footnote{For example, one might expect that the asymptotic form of the Bessel function of Eq.~\ref{eq:besselasymp} could be extended to include more terms in the series expansion, as the Bessel function is an entire function.  However, after inverse Laplace transformation, one can use a ratio test to show that the series expansion for the function $f(q)$ does not converge for $q>-1/2$.}

Then, with Post's approximation for the inverse Laplace transform, this three-point event correlator as a function of distance $d^2$ can be expressed as
\begin{widetext}
\begin{align}
p_{ABC}(d^2)&=\frac{1}{d^4}e^{
3[K_H(\mu,\mu_H)+K_J(\mu,\mu_J)+K_S(\mu,\mu_S)]
}H_{ABC}(\mu_H)J_{ABC}\left(L\to\frac{\partial_{\omega_J}}{3}\right)S_{ABC}\left(L\to\frac{\partial_{\omega_S}}{3}\right)\\
&\hspace{1cm}\times\left(
\frac{\mu_H^2}{E^2R^2}
\right)^{3\omega_H(\mu,\mu_H)}\left(
\frac{\mu_J^2 e^{\gamma_E}}{d^2}
\right)^{3\omega_J(\mu,\mu_J)}\left(
\frac{\mu_S^2E^2R^2e^{2\gamma_E}}{d^4}
\right)^{3\omega_S(\mu,\mu_S)}f\left(
\omega_J(\mu,\mu_J)+2\omega_S(\nu,\mu_s)
\right)\,.
\nonumber
\end{align}
\end{widetext}
Ideally, we would want to take the double-logarithmic limit of this expression, to produce a distribution with a compact, closed functional form to gain some intuition as to its behavior and dependence on parameters.  However, as mentioned earlier, this requires an approximation for the function $f(q)$ near the origin, which in turn requires an approximation of the Bessel function about the index $-1/2$, through Eq.~\ref{eq:laplcefunc3pt}.

\begin{figure}[t!]
\begin{center}
\includegraphics[width=0.45\textwidth]{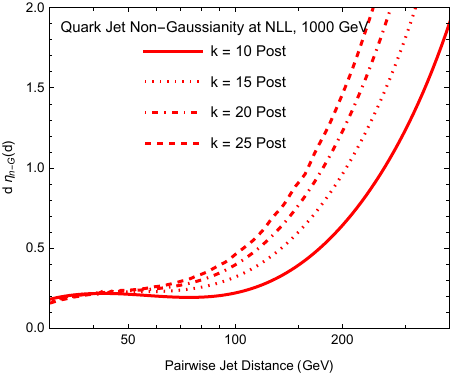}
\caption{\label{fig:qpost}
Distribution of the non-Gaussianity $\eta_\text{n-G}$ as a function of pairwise event distance in GeV, as measured on the 1000 GeV quark jet sample.  Canonical or central renormalization scales are fixed and we show the effect of variation of the accuracy of the approximation to the inverse Laplace transform, with $k=10$ (solid), $k=15$ (dotted), $k=20$ (dot-dashed), or $k=25$ (dashed) derivatives in Post's approximation.
}
\end{center}
\end{figure}

\begin{figure}[t!]
\begin{center}
\includegraphics[width=0.45\textwidth]{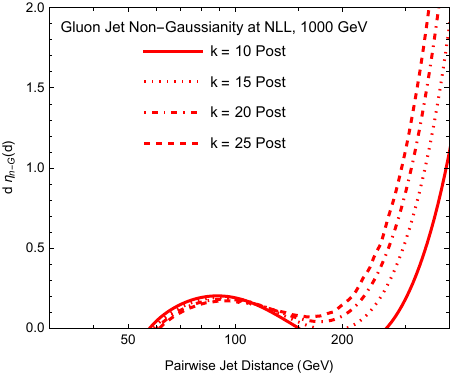}
\caption{\label{fig:gpost}
Distribution of the non-Gaussianity $\eta_\text{n-G}$ as a function of pairwise event distance in GeV, as measured on the 1000 GeV gluon jet sample.  Canonical or central renormalization scales are fixed and we show the effect of variation of the accuracy of the approximation to the inverse Laplace transform, with $k=10$ (solid), $k=15$ (dotted), $k=20$ (dot-dashed), or $k=25$ (dashed) derivatives in Post's approximation.
}
\end{center}
\end{figure}

The implication of this finite-order truncation on the predictions presented in this paper are rather significant.  We have illustrated this in Figs.~\ref{fig:qpost} and \ref{fig:gpost}, for the distribution of the non-Gaussianity as measured on the quark and gluon jets, respectively.  Here, we restrict our attention to jets with energy $1000$ GeV, and plot the distributions for which we use $k=10,15,20,25$ derivatives in Post's approximation for the inverse Laplace transform.  Significant variation is observed as a function of $k$, and further, for gluon jets, significantly high-order derivatives are needed for the distribution to be positive at all.

\section{Explicit Calculation of Two-Loop Anomalous Dimension}\label{app:twoloop}

While we only need one-loop anomalous dimensions to resum to next-to-leading logarithmic accuracy, in this appendix, we will present the calculation of the two-loop anomalous dimension of the two-point event correlator.  This, along with the complete one-loop jet, soft, and hard functions for the jet mass, enables resummation to next-to-next-to-leading logarithmic (NNLL) accuracy.  We leave higher-precision calculations and predictions of complete distributions, however, to future work.

To determine the two-loop anomalous dimension, we focus on the definition of the jet function $J_{AB}(\mu,d^2)$ and its perturbative expansion at two-loop accuracy.  From the definition of the jet function in Eq.~\ref{eq:jetsoftdefs}, the contribution at order-$\alpha_s^2$ takes the form
\begin{align}
&J_{AB}(\mu,d^2)\supset 2\int d\Pi\,  J^{(2)}(\Pi)\,\delta\left(
 d ^2 - s
\right)\\
&\hspace{-0.5cm}+\int d\Pi_A\, d\Pi_B\, J^{(1)}(\Pi_A)\, J^{(1)}(\Pi_B)\,\delta\left(
 d ^2 - \text{SEMD}(\Pi_A,\Pi_B)
\right)\nonumber\,.
\end{align}
Here, $J^{(2)}(\Pi)$ is the two-loop jet function or collinear splitting function on phase space $\Pi$ \cite{Campbell:1997hg,Catani:1999ss,Bern:1998sc,Kosower:1999rx,Bern:1999ry}, and so the first contribution to the two-point event jet function is just twice the jet mass jet function.  On the second line, there is the first mixed contribution, where $J^{(1)}(\Pi)$ is the one-loop jet function or collinear splitting function on phase space $\Pi$.  Now, the metric distance is not simply one jet's invariant mass, so a new calculation must be done in this term.

Given this form of the two-point event correlator, the two-loop non-cusp anomalous dimension of the Laplace-transformed jet function $J_{ab}(\nu)$ takes the form
\begin{align}
\gamma_J^{(2)} = 2\gamma_m^{(2)}+\Delta\gamma_{AB}^{(2)}\,,
\end{align}
where $\gamma_m^{(2)}$ is the two-loop non-cusp anomalous dimension for the jet mass, and $\Delta\gamma_{AB}^{(2)}$ is the anomalous dimension from these non-trivial correlations that we will calculate.  For both quark and gluon jets, the two-loop jet mass anomalous dimensions are well-known \cite{Neubert:2004dd,Becher:2009th}.

For the correlated term, we note that the one-loop jet function on dimensionally-regulated phase space in $D = 4-2\epsilon$ spacetime dimensions is
\begin{align}
&\int d\Pi\, J^{(1)}(\Pi) \\
&\hspace{1cm}\supset \frac{\alpha_s}{2\pi}\frac{\mu^{2\epsilon}}{\Gamma(1-\epsilon)} \int \frac{ds}{s^{1+\epsilon}}\, dz\, z^{-\epsilon}(1-z)^{-\epsilon}\,P(z)\,,\nonumber
\end{align}
where $s$ is the invariant mass of the two particles in the jet, $z$ is the energy fraction of one of the particles, and $P(z)$ is the splitting function.  Note that the integration domain of the invariant mass is $s\in[0,\infty)$, while for the energy fraction it is $z\in[0,1]$.  The relevant cross term with two jet functions is therefore
\begin{widetext}
\begin{align}
&\int d\Pi_A\, d\Pi_B\, J^{(1)}(\Pi_A)\, J^{(1)}(\Pi_B)\, \delta\left(
d^2 - \text{SEMD}(\Pi_A,\Pi_B)
\right) \\
&\hspace{1cm}=  \left(\frac{\alpha_s}{2\pi}\right)^2\frac{\mu^{4\epsilon}}{\Gamma(1-\epsilon)^2} \int \frac{ds_A}{s_A^{1+\epsilon}}\,\frac{ds_B}{s_B^{1+\epsilon}}\, dz_A\, z_A^{-\epsilon}(1-z_A)^{-\epsilon}\, dz_B\, z_B^{-\epsilon}(1-z_B)^{-\epsilon}\,P(z_A)\,P(z_B)\nonumber\\
&\hspace{3cm}\times\delta\left(
d^2-s_A-s_B+2\min[z_A(1-z_A),z_B(1-z_B)]\sqrt{\frac{s_A s_B}{z_Az_B(1-z_A)(1-z_B)}}
\right)\,.
\nonumber
\end{align}
\end{widetext}
We are most interested in the contribution beyond just the sum of the jet masses, as that can be calculated through convolution of the known jet functions for the mass.  So, we will actually consider the difference of this jet function calculation with the sum of the jet masses:
\begin{widetext}
\begin{align}
\Delta J_{AB}(d^2) 
&\equiv  \left(\frac{\alpha_s}{2\pi}\right)^2\frac{\mu^{4\epsilon}}{\Gamma(1-\epsilon)^2} \int \frac{ds_A}{s_A^{1+\epsilon}}\,\frac{ds_B}{s_B^{1+\epsilon}}\, dz_A\, z_A^{-\epsilon}(1-z_A)^{-\epsilon}\, dz_B\, z_B^{-\epsilon}(1-z_B)^{-\epsilon}\,P(z_A)\,P(z_B)\\
&\hspace{-1cm}\times\left[\delta\left(
d^2-s_A-s_B+2\min[z_A(1-z_A),z_B(1-z_B)]\sqrt{\frac{s_A s_B}{z_Az_B(1-z_A)(1-z_B)}}
\right)-\delta\left(
d^2-s_A-s_B
\right)\right]\,.
\nonumber
\end{align}
\end{widetext}

To evaluate this, we will introduce the new variables
\begin{align}
&s \equiv s_A\,, &xs \equiv s_B\,.
\end{align}
We can then integrate over $s$ with the $\delta$-functions to produce
\begin{widetext}
\begin{align}
\Delta J_{AB}(d^2) 
&=  \left(\frac{\alpha_s}{2\pi}\right)^2\frac{\mu^{4\epsilon}}{\Gamma(1-\epsilon)^2} \frac{1}{(d^2)^{1+2\epsilon}}\int\frac{dx}{x^{1+\epsilon}}\, dz_A\, z_A^{-\epsilon}(1-z_A)^{-\epsilon}\, dz_B\, z_B^{-\epsilon}(1-z_B)^{-\epsilon}\,P(z_A)\,P(z_B)\\
&\hspace{1cm}\times\left[\left(1+x-2\min[z_A(1-z_A),z_B(1-z_B)]\sqrt{\frac{x}{z_Az_B(1-z_A)(1-z_B)}}
\right)^{2\epsilon}-\left(1+x\right)^{2\epsilon}\right]\,.\nonumber
\end{align}
\end{widetext}

In the limit in which both jets have a particle that goes soft, there is a residual scaling infrared divergence that must be regulated.  To identify this term, let's consider the energy fractions on the range $z\in[0,1/2]$, and symmetrize the splitting functions appropriately.  That is, there is a single soft divergence in the integrals where $z\to 0$.  To isolate its contribution, we can expand to leading power in the double soft limit $z_A,z_B\to 0$, which produces
\begin{align}
\Delta J_{AB}(d^2) 
&\supset  \left(\frac{\alpha_s}{2\pi}\right)^2\frac{\mu^{4\epsilon}}{\Gamma(1-\epsilon)^2} \frac{C_J^2}{(d^2)^{1+2\epsilon}}\\
&\hspace{-1.5cm}\times\int\frac{dx}{x^{1+\epsilon}}\, \frac{dz_A}{z_A^{1+\epsilon}}\, \frac{dz_B}{z_B^{1+\epsilon}}\,\Theta(1/2-z_A)\Theta(1/2-z_B)\nonumber\\
&\hspace{-1.5cm}\times\left[\left(1+x-2\min[z_A,z_B]\sqrt{\frac{x}{z_Az_B}}
\right)^{2\epsilon}-\left(1+x\right)^{2\epsilon}\right]\,,\nonumber
\end{align}
where $C_J$ is the quadratic color Casimir of the jet.  Now, we can introduce the variables
\begin{align}
&z\equiv z_A\,,&z_B\equiv rz\,,
\end{align}
and so the contribution to the distribution becomes
\begin{align}
\Delta J_{AB}(d^2) 
&\supset  -\left(\frac{\alpha_s}{2\pi}\right)^2\frac{\mu^{4\epsilon}}{\Gamma(1-\epsilon)^2} \frac{2^{2\epsilon}C_J^2}{(d^2)^{1+2\epsilon}}\frac{1}{2\epsilon}\\
&\hspace{-0.5cm}\times\int\frac{dx}{x^{1+\epsilon}}\,  \frac{dr}{r^{1+\epsilon}}\,\max[1,r]^{2\epsilon}\nonumber\\
&\hspace{-0.5cm}\times\left[\left(1+x-2\min[1,r]\sqrt{\frac{x}{r}}
\right)^{2\epsilon}-\left(1+x\right)^{2\epsilon}\right]\nonumber\,.
\end{align}

The integrals that now remain are finite with $\epsilon \to 0$.  We will focus on the divergent term as that introduces novel contributions to the anomalous dimensions at two-loops.  Expanding the integral that remains to lowest order in $\epsilon$, we find
\begin{align}
\Delta J_{AB}(d^2) 
&\supset  \left(\frac{\alpha_s}{2\pi}\right)^2\frac{C_J^2}{\Gamma(1-\epsilon)^2} \frac{(\sqrt{2}\mu)^{4\epsilon}}{(d^2)^{1+2\epsilon}}\\
&\hspace{-0.5cm}\times\int\frac{dx}{x}\,  \frac{dr}{r}\,\log\frac{1+x}{1+x-2\min[1,r]\sqrt{\frac{x}{r}}}\nonumber\\
&\hspace{-0.5cm}=\left(\frac{\alpha_s}{2\pi}\right)^2\frac{C_J^2}{\Gamma(1-\epsilon)^2} \frac{(\sqrt{2}\mu)^{4\epsilon}}{(d^2)^{1+2\epsilon}}\left(
6\pi^2\log 2-7\zeta_3
\right)
\,,\nonumber
\end{align}
where $\zeta_3\approx 1.202$ is the value of the Riemann $\zeta$-function. The correlated contribution to the non-cusp anomalous dimension is therefore
\begin{align}
\hspace{-0.2cm}\Delta\gamma_{AB}^{(2)}= -2 \left(\frac{\alpha_s}{2\pi}\right)^2C_J^2 \left(
6\pi^2\log 2-7\zeta_3
\right)\,.
\end{align}

\section{Two-Loop Anomalous Dimension of Tangent EMD}\label{app:temdrg}

We present calculations for the multi-point event correlators using the SEMD metric because it is simple and fast to evaluate, both for theory calculations as well as on (simulated) data.  However, the first metric for collider event data, and as of now, the most widely-used, is the Energy Mover's Distance (EMD) \cite{Komiske:2019fks}.  For events with fixed total energy $E$, the $\beta$-EMD between two events $A$ and $B$ is defined as
\begin{align}
\text{EMD}({\cal E}_A,{\cal E}_B) = \min_{\{f_{ij}\}}\sum_{i\in A,j\in B} f_{ij}\theta_{ij}^\beta\,,
\end{align}
where $\theta_{ij}$ is the angle between particles in different events, $\beta > 0$ is an angular-weighting parameter, and the coefficient function is constrained as
\begin{align}
f_{ij} \geq 0\,, \sum_{i \in A}f_{ij}\leq E\,, \sum_{j\in B} f_{ij}\leq E\,, \sum_{i\in A,j\in B}f_{ij} = E\,.
\end{align}

In general, the optimization over the energy weight $f_{ij}$ cannot be performed exactly in generality, and further, the EMD is sensitive to event isotropies.  For example, the EMD between two jets that are otherwise identical, but are merely azimuthally-rotated about their respective jet axes from one another, is non-zero.  This is in contrast to the SEMD, which is invariant to such isotropies.  One can, however, define a tangent EMD (TEMD) \cite{pele2013tangent}, which further minimizes the value of the EMD over isometries.  While the (T)EMD cannot be expressed in complete generality, one can solve the optimization if the multiplicity of particles in the events is small enough.

A detailed comparison of the SEMD and the (T)EMD was performed in Ref.~\cite{Larkoski:2023qnv}.  There, it was demonstrated that the $\beta = 2$ TEMD between jets each which have two particles can be expressed as
\begin{align}
&\text{TEMD}^{(1)}_{\beta = 2}(\{s_A,z_A\},\{s_B,z_B\}) \\
&= \min_{\substack{a\in J_A, b\in J_B\\z_a+z_b\leq 1}}\frac{s_A+s_B-2z_az_b\sqrt{\frac{s_As_B}{z_A(1-z_A)z_B(1-z_B)}}}{E}\nonumber\,,
\end{align}
where $s$ is the squared invariant mass of a jet and $z$ is the energy fraction of one particle in the jet.  This takes a very similar form to that of the SEMD on the same type of jets, Eq.~\ref{eq:semdtwopart}, but is in general distinct.  Nevertheless, we will show that the $\beta = 2$ TEMD reduces to the exact same expression as the SEMD as relevant for calculation of the two-loop anomalous dimension of the jet function.  As such, the distributions of the two-point event correlators of the $p=2$ SEMD and $\beta = 2$ TEMD will be identical through next-to-next-to-leading logarithmic accuracy.  

For the anomalous dimension, all that matters is the soft limit of the expression of the TEMD, as demonstrated for the SEMD.  In the soft limit, this reduces to
\begin{align}
&\text{TEMD}^{(1)}_{\beta = 2}(\{s_A,z_A\},\{s_B,z_B\}) \\
&\hspace{1cm}\to \min_{\substack{a\in J_A, b\in J_B\\z_a+z_b\leq 1}}\frac{s_A+s_B-2z_az_b\sqrt{\frac{s_As_B}{z_Az_B}}}{E}\nonumber\,.
\end{align}
In this limit, the minimization constraint becomes simple and reduces to selecting the smaller of $z_A(1-z_B)$ and $z_B(1-z_A)$.  In the soft limit, the TEMD simplifies to
\begin{align}
&\text{TEMD}^{(1)}_{\beta = 2}(\{s_A,z_A\},\{s_B,z_B\}) \\
&\hspace{1cm}\to\frac{s_A+s_B-2\sqrt{s_As_B}\sqrt{\frac{\min[z_A,z_B]}{\max[z_A,z_B]}}}{E}\nonumber\,,
\end{align}
which is identical to the SEMD, up to the rescaling by the total jet energy $E$.  Therefore, their one- and two-loop anomalous dimensions are equal and correspondingly they have identical resummation through NNLL accuracy.  Beyond this order, however, the SEMD and TEMD will in general differ.

\bibliography{refs}

\end{document}